\documentclass{hch}
\usepackage{amssymb}
\usepackage{amsmath}
\usepackage{graphicx}
\usepackage{bm}
\usepackage{amsfonts}

\input{tcilatex}

\begin{document}

\title{Superconducting Qubits and the Physics of Josephson Junctions}
\author{John M. Martinis}

\address{National Institute of Standards and Technology, 325 Broadway,
Boulder, CO 80305-3328, USA}

\authorsup{John M. Martinis\inst[National Institute of Standards and Technology, 325 Broadway, Boulder, CO 80305-3328, USA]{1},
	     Kevin Osborne\inst{1}}

\runningtitle{Martinis \etal: Superconducting Qubits \dots}

\maketitle

\section{\label{intro}Introduction}

\bigskip

Josephson junctions are good candidates for the construction of quantum bits
(qubits) for a quantum computer\cite{QuantumComputing}. \ This system is
attractive because the low dissipation inherent to superconductors make
possible, in principle, long coherence times. \ In addition, because complex
superconducting circuits can be microfabricated using integrated-circuit
processing techniques, scaling to a large number of qubits should be
relatively straightforward. \ Given the initial success of several types of
Josephson qubits\cite%
{Nakamura,NewNakamura,Vion02,Han,Martinis,Chioresku02,NEC03,Maryland03,Simmonds}%
, a question naturally arises: what are the essential components that must
be tested, understood, and improved for eventual construction of a Josephson
quantum computer? \ 

In this paper we focus on the physics of the Josephson junction because,
being nonlinear, it is the fundamental circuit element that is needed for
the appearance of usable qubit states. \ In contrast, \textit{linear}
circuit elements such as capacitors and inductors can form low-dissipation
superconducting resonators, but are unusable for qubits because the
energy-level spacings are degenerate. \ The nonlinearity of the Josephson
inductance breaks the degeneracy of the energy level spacings, allowing
dynamics of the system to be restricted to only the two qubit states. \ The
Josephson junction is a remarkable nonlinear element because it combines
negligible dissipation with extremely large nonlinearity - the change of the
qubit state by only one photon in energy can modify the junction inductance
by order unity!

Most theoretical and experimental investigations with Josephson qubits
assume perfect junction behavior. \ Is such an assumption valid? \ Recent
experiments by our group indicate that coherence is limited by
microwave-frequency fluctuations in the critical current of the junction\cite%
{Simmonds}. \ A deeper understanding of the junction physics is thus needed
so that nonideal behavior can be more readily identified, understood, and
eliminated. \ Although we will not discuss specific imperfections of
junctions in this paper, we want to describe a clear and precise model of
the Josephson junction that can give an intuitive understanding of the
Josephson effect. \ This is especially needed since textbooks do not
typically derive the Josephson effect from a microscopic viewpoint. \ As
standard calculations use only perturbation theory, we will also need to
introduce an exact description of the Josephson effect via the mesoscopic
theory of quasiparticle bound-states.\ \ 

The outline of the paper is as follows. \ We first describe in Sec. \ref%
{nonlinear} the nonlinear Josephson inductance. \ In Sec. \ref{qubittypes}
we discuss the three types of qubit circuits, and show how these circuits
use this nonlinearity in unique manners. \ We then give a brief derivation
of the BCS theory in Sec. \ref{BCS}, highlighting the appearance of the
macroscopic phase parameter. \ The Josephson equations are derived in Sec. %
\ref{Jpert} using standard first and second order perturbation theory that
describe quasiparticle and Cooper-pair tunneling. \ An exact calculation of
the Josephson effect then follows in Sec. \ref{Andreev} using the
quasiparticle bound-state theory. \ Section \ref{QPzener} expands upon this
theory and describes quasiparticle excitations as transitions from the
ground to excited bound states from nonadiabatic changes in the bias. \
Although quasiparticle current is typically calculated only for a constant
DC voltage, the advantage to this approach is seen in Sec. \ref{QPandQubits}%
, where we qualitatively describe quasiparticle tunneling with AC\ voltage
excitations, as appropriate for the qubit state. \ This section describes
how the Josephson qubit is typically insensitive to quasiparticle damping,
even to the extent that a phase qubit can be constructed from microbridge
junctions. \ 

\section{\label{nonlinear}The Nonlinear Josephson Inductance}

\bigskip

A Josephson tunnel junction is formed by separating two superconducting
electrodes with an insulator thin enough so that electrons can
quantum-mechanically tunnel through the barrier, as illustrated in Fig. \ref%
{figJos} . \ The Josephson effect describes the supercurrent $I_{J}$ that
flows through the junction according to the classical equations 
\begin{subequations}
\begin{align}
I_{J}& =I_{0}\sin \delta  \label{Jdc} \\
V& =\frac{\Phi _{0}}{2\pi }\frac{d\delta }{dt}\text{ ,}  \label{Jac}
\end{align}%
where $\Phi _{0}=h/2e$ is the superconducting flux quantum, $I_{0}$ is the
critical-current parameter of the junction, and $\delta =\phi _{L}-\phi _{R}$
and $V$ are respectively the superconducting phase difference and voltage
across the junction. \ The dynamical behavior of these two equations can be
understood by first differentiating Eq.\thinspace \ref{Jdc} and replacing $%
d\delta /dt$ with $V$ according to Eq.\thinspace \ref{Jac} 
\end{subequations}
\begin{equation}
\frac{dI_{J}}{dt}=I_{0}\cos \delta \,\frac{2\pi }{\Phi _{0}}V\text{ .}
\end{equation}%
With $dI_{J}/dt$ proportional to $V$, this equation describes an inductor. \
By defining a Josephson inductance $L_{J}$ according to the conventional
definition $V=L_{J}dI_{J}/dt$, one finds 
\begin{subequations}
\begin{equation}
L_{J}=\frac{\Phi _{0}}{2\pi I_{0}\cos \delta }\text{ .}  \label{LJ}
\end{equation}%
The $1/\cos \delta $ term reveals that this inductance is nonlinear. \ It
becomes large as $\delta \rightarrow \pi /2$, and is negative for $\pi
/2<\delta <3\pi /2$. \ The inductance at zero bias is $L_{J0}=\Phi _{0}/2\pi
I_{0}$.

\FRAME{ftbpFU}{2.5374in}{1.0585in}{0pt}{\Qcb{Schematic diagram of a
Josephson junction connected to a bias voltage $V$. \ The Josephson current
is given by $I_{J}=I_{0}\sin \protect\delta $, where $\protect\delta =%
\protect\phi _{L}-\protect\phi _{R}$ is the difference in the
superconducting phase across the junction.}}{\Qlb{figJos}}{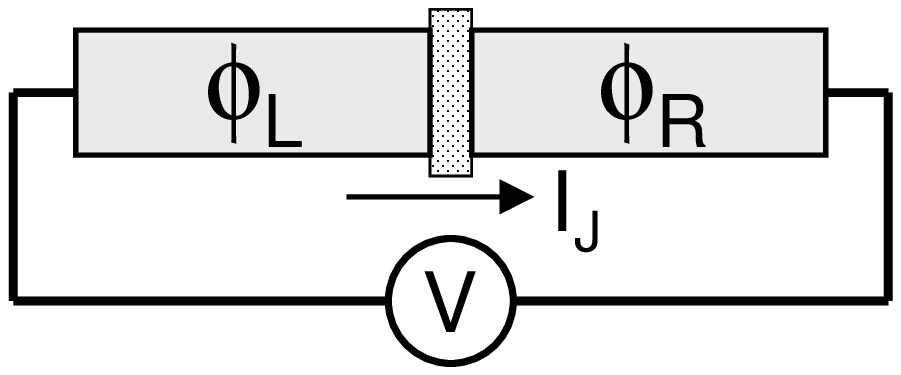}{\special%
{language "Scientific Word";type "GRAPHIC";maintain-aspect-ratio
TRUE;display "USEDEF";valid_file "F";width 2.5374in;height 1.0585in;depth
0pt;original-width 3.5708in;original-height 1.4659in;cropleft "0";croptop
"1";cropright "1";cropbottom "0";filename 'fig1.eps';file-properties
"XNPEU";}}

An inductance describes an energy-conserving circuit element. \ The energy
stored in the junction is given by 
\end{subequations}
\begin{subequations}
\begin{align}
U_{J}& =\dint I_{J}Vdt \\
& =\dint I_{0}\sin \delta \,\frac{\Phi _{0}}{2\pi }\frac{d\delta }{dt}dt \\
& =\frac{I_{0}\Phi _{0}}{2\pi }\dint \sin \delta \,d\delta \\
& =-\frac{I_{0}\Phi _{0}}{2\pi }\cos \delta \text{ .}  \label{EqUj}
\end{align}%
This calculation of energy can be generalized for other nondissipative
circuit elements. \ For example, a similar calculation for a current bias
gives $U_{\mathrm{bias}}=-(I\Phi _{0}/2\pi )\delta $. \ Conversely, if a
circuit element has an energy $U(\delta )$, then the current-phase
relationship of the element, analogous to Eq.\thinspace \ref{Jdc}, is 
\end{subequations}
\begin{equation}
I_{J}(\delta )=\frac{2\pi }{\Phi _{0}}\frac{\partial U(\delta )}{\partial
\delta }\text{ .}  \label{UtoI}
\end{equation}%
A generalized Josephson inductance can be also be found from the second
derivative of $U$ ,%
\begin{equation}
\frac{1}{L_{J}}=\left( \frac{2\pi }{\Phi _{0}}\right) ^{2}\frac{\partial
^{2}U(\delta )}{\partial \delta ^{2}}\text{ .}
\end{equation}

The classical and quantum behavior of a particular circuit is described by a
Hamiltonian, which of course depends on the exact circuit configuration. \
The procedure for writing down a Hamiltonian for an arbitrary circuit has
been described in detail in a prior publication\cite{MichelLesHouches}. The
general form of the Hamiltonian for the Josephson effect is $H_{J}=U_{J}$. \
\ 

\section{\label{qubittypes}Phase, Flux, and Charge Qubits}

\bigskip

A Josephson qubit can be understood as a nonlinear resonator formed from the
Josephson inductance and its junction capacitance. \ nonlinearity is crucial
because the system has many energy levels, but the operating space of the
qubit must be restricted to only the two lowest states. \ The system is
effectively a two-state system\cite{Steffen03} only if the frequency $\omega
_{10}$ that drives transitions between the qubit states $%
0\longleftrightarrow 1$ is different from the frequency $\omega _{21}$ for
transitions $1\longleftrightarrow 2$. \ 

We review here three different ways that these nonlinear resonators can be
made, and which are named as phase, flux, or charge qubits. \ \ 

The circuit for the phase-qubit circuit is drawn in Fig. \ref{figcompare}%
(a). \ Its Hamiltonian is%
\begin{equation}
H=\frac{1}{2C}\widehat{Q}^{2}-\frac{I_{0}\Phi _{0}}{2\pi }\cos \widehat{%
\delta }-\frac{I\Phi _{0}}{2\pi }\widehat{\delta }\text{ ,}  \label{Hphase}
\end{equation}%
where $C$ is the capacitance of the tunnel junction. \ A similar circuit is
drawn for the flux-qubit circuit in Fig. \ref{figcompare}(b), and its
Hamiltonian is 
\begin{equation}
H=\frac{1}{2C}\widehat{Q}^{2}-\frac{I_{0}\Phi _{0}}{2\pi }\cos \widehat{%
\delta }+\frac{1}{2L}(\Phi -\frac{\Phi _{0}}{2\pi }\widehat{\delta })^{2}%
\text{ \ .}
\end{equation}%
The charge qubit has a Hamiltonian similar to that in Eq. \ref{Hphase}, and
is described elsewhere in this publication. \ Here we have explicitly used
notation appropriate for a quantum description, with operators charge $%
\widehat{Q}$ and phase difference $\widehat{\delta }$ that obey a
commutation relationship $[\widehat{\delta },\widehat{Q}]=2ei$. \ Note that
the phase and flux qubit Hamiltonians are equivalent for $L\rightarrow
\infty $ and $I=\Phi /L$, which corresponds to a current bias created from
an inductor with infinite impedance.

\FRAME{ftbpFU}{3.5405in}{2.0487in}{0pt}{\Qcb{Comparison of the phase (a),
flux (b), and charge (c) qubits. \ Top row illustrates the circuits, with
each \textquotedblleft X\textquotedblright\ symbol representing a Josephson
juncton. \ Middle row has a plot of the Hamiltonian potential (thick line),
showing qualitatively different shapes for three qubit types. \ Ground-state
wavefunction is also indicated (thin line). \ \ Key circuit parameters are
listed in next row. \ Lowest row indicates variations on the basic circuit,
as discussed in text. \ The lowest three energy levels are illustrated for
the phase qubit (dotted lines). \ }}{\Qlb{figcompare}}{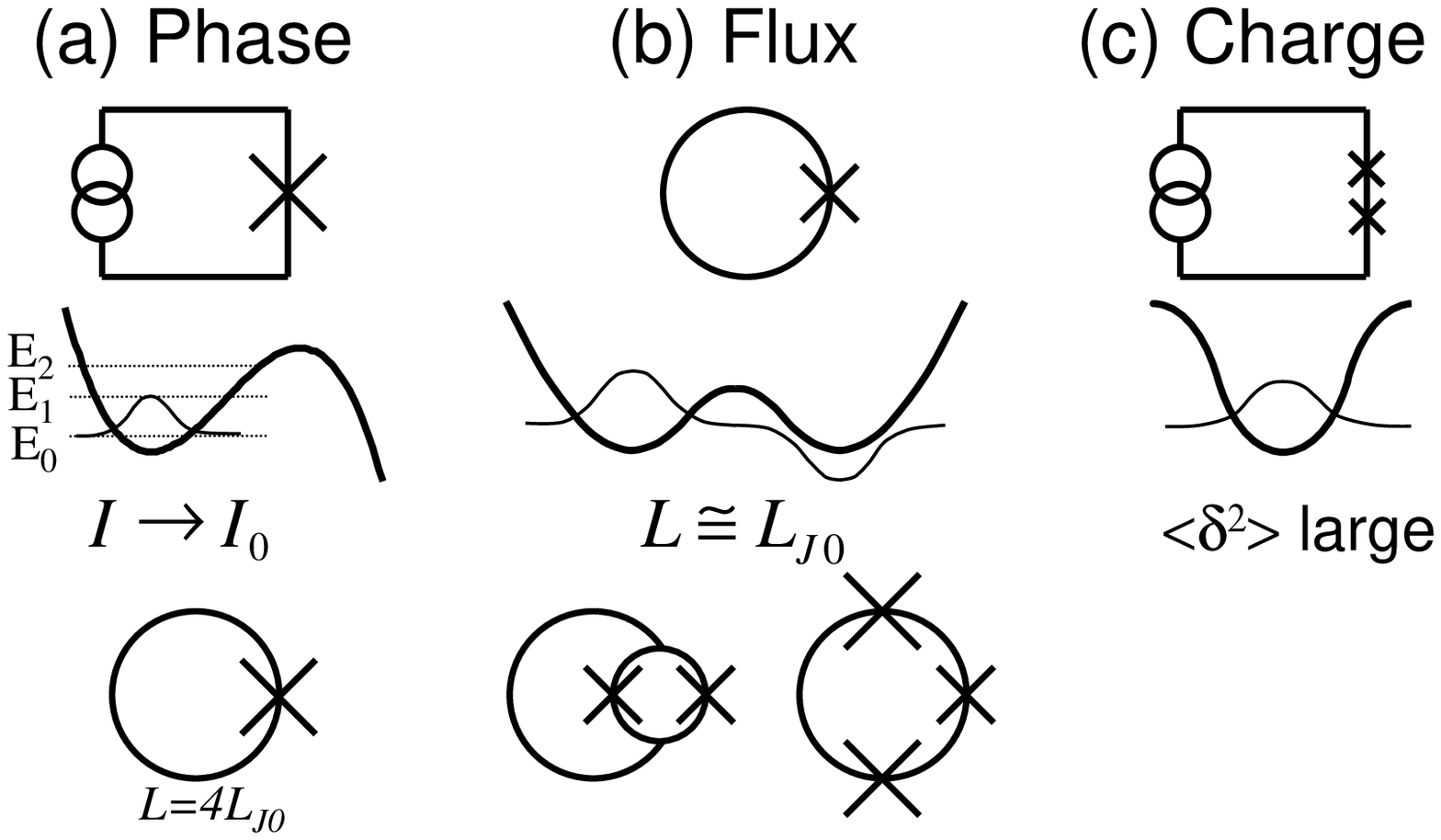}{\special%
{language "Scientific Word";type "GRAPHIC";maintain-aspect-ratio
TRUE;display "USEDEF";valid_file "F";width 3.5405in;height 2.0487in;depth
0pt;original-width 6.4316in;original-height 3.6997in;cropleft "0";croptop
"1";cropright "1";cropbottom "0";filename 'fig2.eps';file-properties
"XNPEU";}}The commutation relationship between $\widehat{\delta }$ and $%
\widehat{Q}$ imply that these quantities must be described by a
wavefunction. \ The characteristic widths of this wavefunction are
controlled by the energy scales of the system, the charging energy of the
junction $E_{C}=e^{2}/2C$ and the Josephson energy $E_{J}=I_{0}\Phi
_{0}/2\pi $. \ When the energy of the junction dominates, $E_{J}\gg E_{C}$,
then $\widehat{\delta }$ can almost be described classically and the width
of its wavefunction is small $\langle \widehat{\delta }^{2}-\langle \widehat{%
\delta }\rangle ^{2}\rangle \ll 1$. \ In contrast, the uncertainty in charge
is large $\langle \widehat{Q}^{2}-\langle \widehat{Q}\rangle ^{2}\rangle \gg
(2e)^{2}$. \ 

If the Josephson inductance is constant over the width of the $\widehat{%
\delta }$ wavefunction, then a circuit is well described as a $L_{J}$-$C$
harmonic oscillator, and the qubit states are degenerate and not usable. \
Usable states are created only when the Josephson inductance changes over
the $\delta $-wavefunction.

The most straightforward way for the wavefunction to be affected by the
Josephson nonlinearity is for $\widehat{\delta }$ to have a large width ,
which occurs when $E_{J}\sim E_{C}$. \ A practical implementation of this
circuit is illustrated in Fig \ref{figcompare}(c), where a double-junction
Coulomb blockade device is used instead of a single junction to isolate
dissipation from the leads\cite{Nakamura,Vion02}. \ Because the wavefunction
extends over most of the -$\cos \widehat{\delta }$ Hamiltonian, the
transition frequency $\omega _{10}$ can differ from $\omega _{21}$ by more
than $10\,\%$, creating usable qubit states\cite{CottetThesis}. \ 

Josephson qubits are possible even when $E_{J}\gg E_{C}$, provided that the
junction is biased to take advantage of its strong nonlinearity. \ A good
example is the phase qubit\cite{Martinis}, where typically $E_{J}\sim
10^{4}E_{C}$, but which is biased near $\delta \lessapprox \pi /2$ so that
the inductance changes rapidly with $\delta $ (see Eq.\thinspace \ref{LJ}).
\ Under these conditions the potential can be accurately described by a
cubic potential, with the barrier height $\Delta U\rightarrow 0$ as $%
I\rightarrow I_{0}$. \ Typically the bias current is adjusted so that the
number of energy levels in the well is $\sim 3-5$, which causes $\omega
_{10} $ to differ from $\omega _{21}$ by an acceptably large amount $\sim
5\,\%$. \ \ 

Implementing the phase qubit is challenging because a current bias is
required with large impedance. \ This impedance requirement can be met by
biasing the junction with flux through a superconducting loop with a large
loop inductance $L$, as discussed previously and drawn in Fig. \ref%
{figcompare}(a). \ To form multiple stable flux states and a cubic
potential, the loop inductance $L$ must be chosen such that $L\gtrsim
2L_{J0} $. \ We have found that a design with $L\simeq 4.5L_{J0}$ is a good
choice since the potential well then contains the desired cubic potential
and only one flux state into which the system can tunnel, simplifying
operation. \ 

The flux qubit is designed with $L\lessapprox L_{J0}$ and biased in flux so
that $\langle \widehat{\delta }\rangle =\pi $. \ Under these conditions the
Josephson inductance is negative and is almost canceled out by $L$. \ The
small net negative inductance near $\widehat{\delta }=\pi $ turns positive
away from this value because of the $1/\cos \delta $ nonlinearity, so that
the final potential shape is quartic, as shown in Fig. \ref{figcompare}(b).
\ An advantage of the flux qubit is a large net nonlinearity, so that $%
\omega _{10}$ can differ from $\omega _{21}$ by over $100\,\%$. \ 

The need to closely tune $L$ with $L_{J0}$ has inspired the invention of
several variations to the simple flux-qubit circuit, as illustrated in Fig. %
\ref{figcompare}(b). \ One method is to use small area junctions\cite%
{Chioresku02} with $E_{J}\sim 10E_{C}$, producing a large width in the $%
\widehat{\delta }$ wavefunction and relaxing the requirement of close tuning
of $L$ with $L_{J0}$. \ Another method is to make the qubit junction a
two-junction SQUID, whose critical current can then be tuned via a second
flux-bias circuit\cite{SUNYrf,KochRF}. \ Larger junctions are then
permissible, with $E_{J}\sim 10^{2}E_{C}$ to $10^{3}E_{C}$. \ A third method
is to fabricate the loop inductance from two or more larger critical-current
junctions\cite{Delft3j}. \ These junctions are biased with phase less than $%
\pi /2$, and thus act as positive inductors. \ The advantage to this
approach is that junction inductors are smaller than physical inductors, and
fabrication imperfections in the critical currents of the junctions tend to
cancel out and make the tuning of $L$ with $L_{J0}$ easier.

In summary, the major difference between the phase, flux, and charge qubits
is the shape of their nonlinear potentials, which are respectively cubic,
quartic, and cosine. \ It is impossible at this time to predict which qubit
type is best because their limitations are not precisely known, especially
concerning decoherence mechanisms and their scaling. \ However, some general
observations can be made.

First, the flux qubit has the largest nonlinearity. \ This implies faster
logic gates since suppressing transitions from the qubit states $0$ and $1$
to state $2$ requires long pulses whose time duration scales as $%
1/\left\vert \omega _{10}-\omega _{21}\right\vert $\cite{Steffen03}. \ The
flux qubit allows operation times less than $\sim 1\,\mathrm{ns}$, whereas
for the phase qubit $10\,\mathrm{ns}$ is more typical. \ We note, however,
that this increase in speed may not be usable. \ Generating precise shaped
pulses is much more difficult on a $1\,\mathrm{ns}$ time scale, and
transmitting these short pulses to the qubit with high fidelity will be more
problematic due to reflections or other imperfections in the microwave
lines. \ \ 

Second, the choice between large and small junctions involve tradeoffs. \
Large junctions ($E_{J}$ $\gg E_{C}$) require precise tuning of parameters ($%
L/L_{J0}$ for the flux qubit) or biases ($I/I_{0}$ for the phase qubit) to
produce the required nonlinearity. \ Small junctions ($E_{J}$ $\sim E_{C}$)
do not require such careful tuning, but become sensitve to $1/f$ charge
fluctuations because $E_{C}$ has relatively larger magnitude. \ 

Along these lines, the coherence of qubits have been compared considering
the effect of low-frequency $1/f$ fluctuations of the critical current\cite%
{Clarke1f}. \ These calculations include the known scaling of the
fluctuations with junction size and the sensitivity to parameter
fluctuations. \ It is interesting that the calculated coherence times for
the flux and phase qubits are similar. \ With parameters choosen to give an
oscillation frequency of $\sim 1\,\mathrm{GHz}$ for the flux qubit and $\sim
10\,\mathrm{GHz}$ for the phase qubit, the number of coherent logic-gate
operations is even approximately the same. \ \ 

\section{\label{BCS}BCS Theory and the Superconducting State}

\bigskip

A more complete understanding of the Josephson effect will require a
derivation of Eqs. \ref{Jdc} and \ref{Jac}. \ In order to calculate this
microscopically, we will first review the BCS theory of superconductivity%
\cite{BCS} using a \textquotedblleft pair spin\textquotedblright\ derivation
that we believe is more physically clear than the standard
energy-variational method. \ Although the calculation follows closely that
of Anderson\cite{Anderson} and Kittel\cite{Kittel}, we have expanded it
slightly to describe the physics of the superconducting phase, as
appropriate for understanding Josephson qubits. \ 

In a conventional superconductor, the attractive interaction that produces
superconductivity comes from the scattering of electrons and phonons. \ As
illustrated in Fig. \ref{figphonon}(a), to first order the phonon
interaction scatters an electron from one momentum state to another. \ When
taken to second order (Fig. \ref{figphonon}(b)), the scattering of a virtual
phonon produces a net attractive interaction between two pairs of electrons.
\ The first-order phonon scattering rates are generally small, not because
of the phonon matrix element, but because phase space is small for the final
electron state. \ This implies that the energy of the second order
interaction can be significant if there are large phase-space factors.

\FRAME{ftbpFU}{3.039in}{1.1874in}{0pt}{\Qcb{Feynman diagram of
electron-phonon interaction showing (a) first- and (b) second-order
processes. }}{\Qlb{figphonon}}{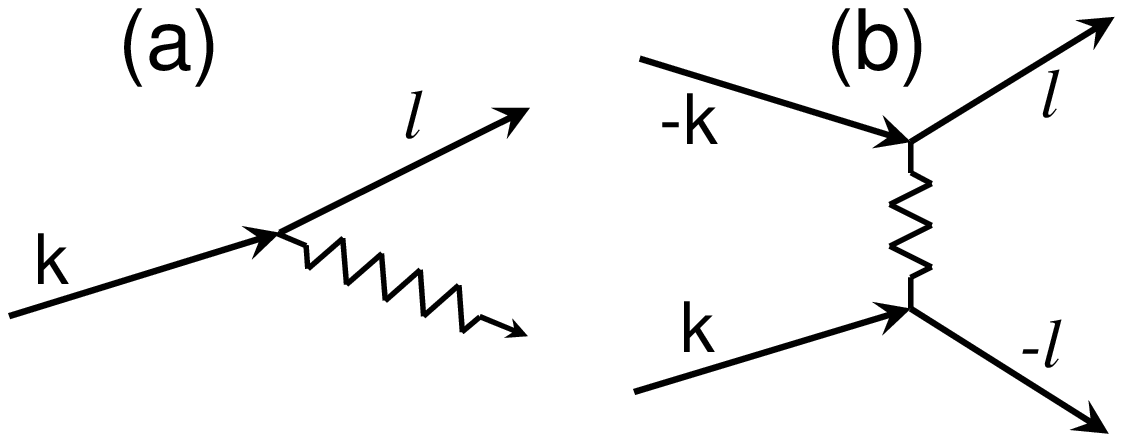}{\special{language "Scientific
Word";type "GRAPHIC";maintain-aspect-ratio TRUE;display "USEDEF";valid_file
"F";width 3.039in;height 1.1874in;depth 0pt;original-width
4.4434in;original-height 1.7115in;cropleft "0";croptop "1";cropright
"1";cropbottom "0";filename 'fig3.eps';file-properties "XNPEU";}}The
electron pairs have the largest net interaction if every pair is allowed by
phase space factors to interact with every other pair. \ This is explicitly
created in the BCS wavefunction by including only pair states (Cooper pairs)
with zero net momentum. \ Under this assumption and using a second quantized
notation where $c_{k}^{\dag }$ is the usual creation operator for an
electron state of wavevector $k$, the most general form for the electronic
wavefunction is%
\begin{equation}
\Psi =\dprod\limits_{k}(u_{k}+v_{k}e^{i\phi _{k}}c_{k}^{\dag }c_{-k}^{\dag
})\left\vert 0\right\rangle \text{ ,}
\end{equation}%
where $u_{k}$ and $v_{k}$ are real and correspond respectively to the
probability amplitude for a pair state to be empty or filled, and are
normalized by $u_{k}^{2}+$ $v_{k}^{2}=1$. \ For generality we have included
a separate phase factor $\phi _{k}$ for each pair. \ Because each pair state
is described as a two state system, the wavefunction may also be described
equivalently with a \textquotedblleft pair-spin\textquotedblright\ tensor
product%
\begin{equation}
\Psi =\dprod\limits_{k}\binom{u_{k}}{v_{k}e^{i\phi _{k}}}\otimes \text{ ,}
\end{equation}%
and the Hamiltonian given with Pauli matrices $\sigma _{xk}$, $\sigma _{yk}$%
, and $\sigma _{zk}$.

\ The kinetic part of the Hamiltonian must give $\Psi $ in the ground state
with pairs occupied only for $\left\vert k\right\vert <k_{f}$, where $k_{f}$%
\ is the Fermi momentum. \ \ If we define the kinetic energy of a single
electron, relative to the Fermi energy, as $\xi _{k}$, then the kinetic
Hamiltonian for the pair state is 
\begin{equation}
H_{K}=-\tsum \xi _{k}\sigma _{zk}\text{ .}
\end{equation}%
The solution of $H_{K}\Psi =E_{k\pm }\Psi $ gives for the lowest energy, $%
E_{k-}$, the values $v_{k}=1$ for $\left\vert k\right\vert <k_{f}$, and $%
v_{k}=0$ for $\left\vert k\right\vert >k_{f}$, as required. \ An energy $%
E_{k+}-E_{k-}=2\left\vert \xi _{k}\right\vert $ is needed for the excitation
of pairs above the Fermi energy or the excitation of holes\ (removal of
pairs) below the Fermi energy.\ 

The potential part of the pair-spin Hamiltonian comes from the second-order
phonon interaction that both creates and destroys a pair, as illustrated in
Fig. \ref{figphonon}(b). \ The Hamiltonian for this interaction is given by 
\begin{equation}
H_{\Delta }=-\frac{V}{2}\dsum\limits_{k,l}(\sigma _{xk}\sigma _{xl}+\sigma
_{yk}\sigma _{yl})\text{ ,}  \label{Hpair}
\end{equation}%
and can be checked to correspond to the second-quantization Hamiltonian $%
H_{\Delta }=-V\tsum c_{k}^{\dag }c_{-k}^{\dag }c_{k}c_{-k}$ by using the
translation $\sigma _{xk}\rightarrow $ $c_{k}c_{-k}+c_{k}^{\dag
}c_{-k}^{\dag }$ and $\sigma _{yk}\rightarrow i(c_{k}c_{-k}-c_{k}^{\dag
}c_{-k}^{\dag })$.\ 

We will first understand the solution to the Hamiltonian $H_{K}+H_{\Delta }$
for the phase variables $\phi _{k}$. \ This Hamiltonian describes a bath of
spins that are all coupled to each other in the x-y plane ($H_{\Delta }$)
and have a distribution of magnetic fields in the z-direction ($H_{K}$). \
Because $H_{\Delta }$ is negative, each pair of spins becomes aligned with
each other in the x-y plane, which implies that every spin in the bath has
the same phase $\phi _{k}$. \ This condition explains why the BCS
wavefunction has only one phase $\phi =\phi _{k}$\ for all Cooper pairs\cite%
{Macrophase}. \ Because there is no preferred direction in the x-y plane,
the solution to the Hamiltonian is degenerate with respect to $\phi $ and
the wavefunction for $\phi $\ is separable from the rest of the
wavefunction. \ Normally, this means that $\phi $\ can be treated as a
classical variable, as is done for the conventional understanding of
superconductivity and the Josephson effects. \ For Josephson qubits, where $%
\phi $ must be treated quantum mechanically, then the behavior of $\phi $ is
described by an external-circuit Hamiltonian, as was done in Sec. \ref%
{qubittypes}.

For a superconducting circuit, where one electrode is biased with a voltage $%
V$, the voltage can be accounted for with a gauge transformation on each
electron state $c_{k}^{\dag }\rightarrow e^{i(e/\hbar )\int Vdt}c_{k}^{\dag
} $. \ The change in the superconducting state is thus given by 
\begin{eqnarray}
\Psi &\rightarrow &\dprod\limits_{k}(u_{k}+v_{k}e^{i\phi }e^{i(e/\hbar )\int
Vdt}c_{k}^{\dag }e^{i(e/\hbar )\int Vdt}c_{-k}^{\dag })\left\vert
0\right\rangle \\
&=&\dprod\limits_{k}(u_{k}+v_{k}e^{i[\phi +i(2e/\hbar )\int Vdt]}c_{k}^{\dag
}c_{-k}^{\dag })\left\vert 0\right\rangle \text{ .}
\end{eqnarray}%
The change in $\phi $ can be written equivalently as 
\begin{equation}
\frac{d\phi }{dt}=\frac{2eV}{\hbar }\text{ ,}
\end{equation}%
which leads to the AC Josephson effect.

The solution for $u_{k}$ and $v_{k}$ proceeds using the standard method of
mean-field theory, with%
\begin{align}
\left\langle H_{\Delta }\right\rangle & =-\frac{V}{2}\dsum\limits_{k,l}(%
\sigma _{xk}\left\langle \sigma _{xl}\right\rangle +\sigma _{yk}\left\langle
\sigma _{yl}\right\rangle )\text{ ,} \\
\left\langle \sigma _{xl}\right\rangle & =\left( u_{l},v_{l}e^{-i\phi
}\right) \cdot \sigma _{x}\cdot \binom{u_{l}}{v_{l}e^{i\phi }} \\
& =2u_{l}v_{l}\cos \phi \text{ ,} \\
\left\langle \sigma _{yl}\right\rangle & =2u_{l}v_{l}\sin \phi \text{ .}
\end{align}%
Using the standard definition of the gap potential, one finds 
\begin{subequations}
\begin{align}
\Delta & =V\dsum\limits_{l}u_{l}v_{l}\text{ ,}  \label{Gapeq} \\
H& =H_{K}+\left\langle H_{\Delta }\right\rangle \\
& =-\dsum\limits_{k}(\sigma _{xk},\sigma _{yk},\sigma _{zk})\cdot (\Delta
\cos \phi ,\Delta \sin \phi ,\xi _{k})\text{ .}
\end{align}

\bigskip \FRAME{ftbpFU}{3.039in}{1.5342in}{0pt}{\Qcb{Bloch sphere solution
of the Hamiltonian $(\protect\sigma _{x},\protect\sigma _{y},\protect\sigma %
_{z})\bullet (B_{x},B_{y},B_{z})$. \ The vector $\protect\overrightarrow{B}$
gives the direction of the positive energy eigenstate. \ }}{\Qlb{figbloch}}{%
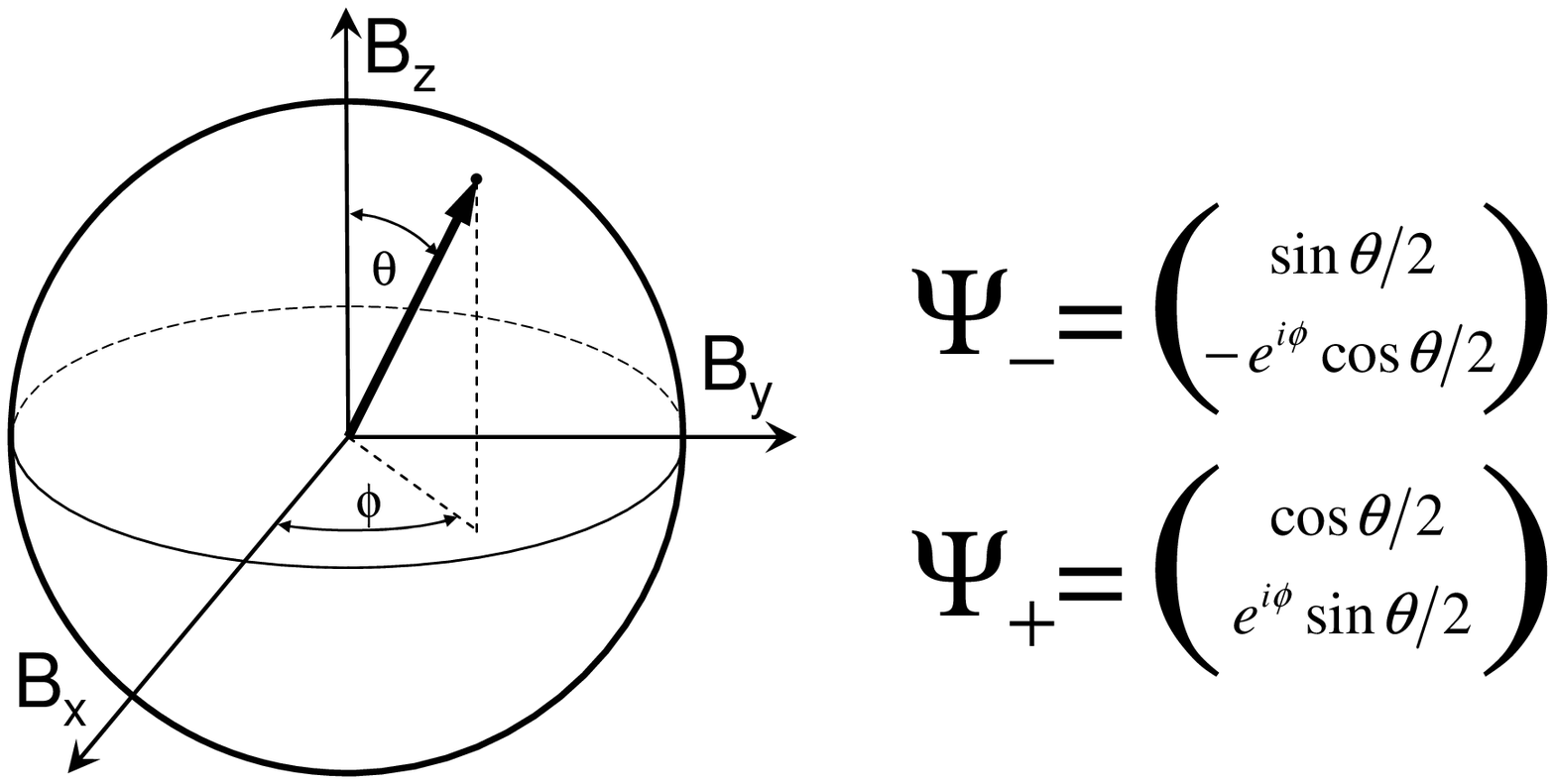}{\special{language "Scientific Word";type
"GRAPHIC";maintain-aspect-ratio TRUE;display "USEDEF";valid_file "F";width
3.039in;height 1.5342in;depth 0pt;original-width 7.3613in;original-height
3.544in;cropleft "0";croptop "1";cropright "1";cropbottom "0";filename
'fig4.eps';file-properties "XNPEU";}}

This Hamiltonian is equivalent to a spin 1/2 particle in a magnetic field,
and its solution is well known. \ The energy eigenvalues of $H\Psi =E_{k\pm
}\Psi $ are given by the total length of the field vector, 
\end{subequations}
\begin{equation}
E_{k\pm }=\pm (\Delta ^{2}+\xi _{k}^{2})^{1/2}\text{ ,}
\end{equation}%
and the directions of the Bloch vectors describing the $E_{k+}$ and $E_{k-}$
eigenstates are respectively parallel and antiparallel to the direction of
the field vector, as illustrated in Fig. \ref{figbloch}. \ The ground state
solution $\Psi _{k-}$ is given by 
\begin{align}
u_{k}& =\sqrt{\frac{1}{2}\left( 1+\frac{\xi _{k}}{E_{k}}\right) }\text{ ,}
\label{equ} \\
v_{k}& =\sqrt{\frac{1}{2}\left( 1-\frac{\xi _{k}}{E_{k}}\right) }\text{ ,}
\label{eqv} \\
\phi _{k}& =\phi \text{ ,}
\end{align}%
with the last equation required for consistency. \ The excited state $\Psi
_{k+}$ is similarly described, but with $u_{k}$ and $v_{k}$ interchanged and 
$\phi \rightarrow \phi +\pi $. \ 

At temperature $T=0$ the energy gap $\Delta $ may be solved by inserting the
solutions for $u_{k}$ and $v_{k}$ into Eq. \ref{Gapeq}%
\begin{equation}
\Delta =V\dsum\limits_{l}\frac{\Delta }{2(\Delta ^{2}+\xi _{k}^{2})^{1/2}}%
\text{ .}
\end{equation}%
Converting to an integral by defining a density of states $N_{0}$ at the
Fermi energy, and introducing a cutoff of the interaction $V$ at the Debye
energy $\theta _{D}$, one finds the standard BCS result, 
\begin{equation}
\Delta =2\theta _{D}e^{-1/N_{0}V}\text{ .}
\end{equation}

Two eigenstates $E_{k-}$ and $E_{k+}$ have been determined for the pair
Hamiltonian. \ Two additional \textquotedblleft
quasiparticle\textquotedblright\ eigenstates must exist, which clearly have
to be single-particle states. \ These states may be solved for using
diagonalization techniques, giving 
\begin{subequations}
\begin{eqnarray}
\Psi _{k0} &=&c_{k}^{\dag }\left\vert 0\right\rangle \text{ ,}
\label{QPstate0} \\
\Psi _{k1} &=&c_{-k}^{\dag }\left\vert 0\right\rangle \text{ .}
\label{QPstate1}
\end{eqnarray}%
Fortunately, these states may be easily checked by inspection. \ The kinetic
part of the Hamiltonian gives $H_{K}\Psi _{k0,1}=0$ since $\Psi _{k0,1}$
corresponds to the creation of an electron and a hole, and the electron-pair
and hole-pair states have opposite kinetic energy. \ The potential part of
the energy also gives $\left\langle H_{\Delta }\right\rangle \Psi _{k0,1}=0$
since the interaction Hamiltonian scatters pair states. \ Thus the
eigenenergies of $\Psi _{k0,1}$ are zero, and these states have an energy $%
E_{k}=\left\vert E_{k-}\right\vert $ above the ground state. \ 

The quasiparticle operators that take the ground-state wavefunction to the
excited states are 
\end{subequations}
\begin{align}
\gamma _{k0}^{\dagger }& =u_{k}c_{k}^{\dag }-v_{k}e^{-i\phi }c_{-k}\text{ ,}
\label{gamma0} \\
\gamma _{k1}^{\dagger }& =u_{k}c_{-k}^{\dag }+v_{k}e^{-i\phi }c_{k}\text{ ,}
\label{gamma1}
\end{align}%
which can be easily checked to give%
\begin{eqnarray}
\gamma _{k0}^{\dagger }(u_{k}+v_{k}e^{i\phi }c_{k}^{\dag }c_{-k}^{\dag
})\left\vert 0\right\rangle &=&c_{k}^{\dag }\left\vert 0\right\rangle \text{
,}  \label{cgamma0} \\
\gamma _{k1}^{\dagger }(u_{k}+v_{k}e^{i\phi }c_{k}^{\dag }c_{-k}^{\dag
})\left\vert 0\right\rangle &=&c_{-k}^{\dag }\left\vert 0\right\rangle \text{
.}  \label{cgamma1}
\end{eqnarray}

A summary of these results is illustrated in Fig. \ref{figgamma}, where we
show the energy levels, wavefunctions, and operators for transitions between
the four states. \ The quasiparticle raising and lowering operators $\gamma
_{k0}^{\dagger }$, $\gamma _{k1}^{\dagger }$, $\gamma _{k0}^{{}}$, and $%
\gamma _{k1}^{{}}$ produce transitions between the states and have
orthogonality relationships similar to those of the electron operators. \ 

\FRAME{ftbpFU}{3.039in}{1.8317in}{0pt}{\Qcb{Energy-level diagram for the
ground-pair state (solid line), two quasiparticle states (dashed lines), and
the excited-pair state (short dashed line). \ }}{\Qlb{figgamma}}{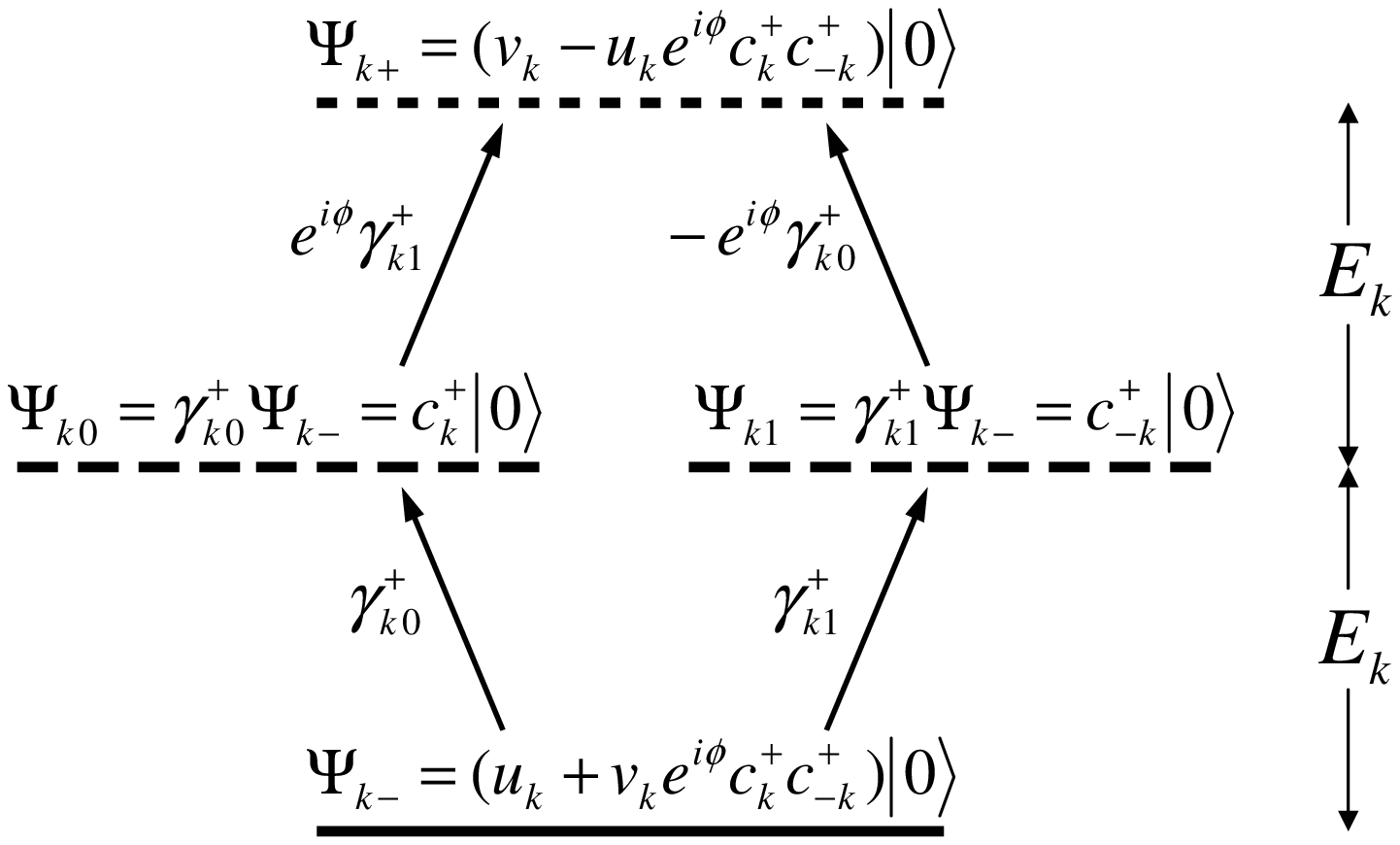}{%
\special{language "Scientific Word";type "GRAPHIC";maintain-aspect-ratio
TRUE;display "USEDEF";valid_file "F";width 3.039in;height 1.8317in;depth
0pt;original-width 5.7398in;original-height 3.4385in;cropleft "0";croptop
"1";cropright "1";cropbottom "0";filename 'fig4a.eps';file-properties
"XNPEU";}}

It is interesting to note that the ground and excited pair states are
connected by the two quasiparticle operators $e^{-i\phi }\gamma
_{k1}^{\dagger }\gamma _{k0}^{\dagger }\Psi _{k-}=\Psi _{k+}$. \ Because the
value $u_{l}v_{l}$ changes sign between $\Psi _{k-}$ and $\Psi _{k+}$, and
is zero for $\Psi _{k0,1}$, the gap equation \ref{Gapeq} including the
effect of quasiparticles is proportional to $\langle 1-\gamma _{k0}^{\dagger
}\gamma _{k0}^{{}}-\gamma _{k1}^{\dagger }\gamma _{k1}^{{}}\rangle $. \
Along with the energy levels, these results imply that the two types of
quasiparticles are independent excitations. \ \ 

\section{\label{Jpert}The Josephson Effect, Derived from Perturbation Theory}

\ 

We will now calculate the quasiparticle and Josephson current for a tunnel
junction using first and second order perturbation theory, respectively. \
We note that our prior calculations have not been concerned with electrical
transport. \ In fact, the electron operators describing the superconducting
state have not been influenced by charge, and thus they correspond to the
occupation of an effectively neutral state. \ Because a tunneling event
involves a real transfer of an electron, charge must now be accounted for
properly. \ We will continue to use electron operators for describing the
states, but will keep track of the charge transfer separately. \ 

When an electron tunnels through the barrier, an electron and hole state is
created on the opposite (left and right) side of the barrier. \ The
tunneling Hamiltonian for this process can be written as 
\begin{subequations}
\begin{eqnarray}
H_{T} &=&\overrightarrow{H}_{T+}+\overrightarrow{H}_{T-}+\overleftarrow{H}%
_{T+}+\overleftarrow{H}_{T-}  \label{HT} \\
&=&\dsum\limits_{L,R}\left( t_{LR}^{{}}c_{L}^{{}}c_{R}^{\dag
}+t_{-L-R}^{{}}c_{-L}^{{}}c_{-R}^{\dag }+t_{LR}^{\ast }c_{L}^{\dagger
}c_{R}^{{}}+t_{-L-R}^{\ast }c_{-L}^{\dagger }c_{-R}^{{}}\right) \text{ ,}
\label{HTterms}
\end{eqnarray}%
where $t_{LR}^{{}}$ is the tunneling matrix element, and the $L$ and $R$
indices refer respectively to momentum states $k$ on the left and right
superconductor. \ The first two terms $\overrightarrow{H}_{T+}$ and $%
\overrightarrow{H}_{T-}$ correspond to the tunneling of one electron from
left to the right, whereas $\overleftarrow{H}_{T+}$ and $\overleftarrow{H}%
_{T-}$ are for tunneling to the left. \ The Hamiltonian is explicitly broken
up into $\overrightarrow{H}_{T+}$ and $\overrightarrow{H}_{T-}$ to account
for the different electron operators $c_{k}^{\dag }$ and $c_{-k}^{\dag }$
for positive and negative momentum. \ \ 

The electron operators must first be expressed in terms of the quasiparticle
operators $\gamma $ because these produce transitions between eigenstates of
the superconducting Hamiltonian. \ Equations \ref{gamma0}, \ref{gamma1}, and
their adjoints are used to solve for the four electron operators 
\end{subequations}
\begin{equation}
\begin{array}{cc}
c_{k}=u_{k}\gamma _{k0}+v_{k}e^{i\phi }\gamma _{k1}^{\dagger } & 
c_{-k}=u_{k}\gamma _{k1}-v_{k}e^{i\phi }\gamma _{k0}^{\dagger } \\ 
c_{k}^{\dagger }=u_{k}\gamma _{k0}^{\dagger }+v_{k}e^{-i\phi }\gamma _{k1} & 
c_{-k}^{\dagger }=u_{k}\gamma _{k1}^{\dagger }-v_{k}e^{-i\phi }\gamma _{k0}%
\text{ .}%
\end{array}
\label{ceqs}
\end{equation}

Substituting Eqs. \ref{ceqs} into \ref{HTterms}, one sees that all four
terms of the Hamiltonian have operators $\gamma ^{\dagger }$ that produce
quasiparticles. \ We calculate here to first order the quasiparticle current
from $L$ to $R$ given by $\overrightarrow{H}_{T+}+\overrightarrow{H}_{T-}$.
\ The Feynman diagrams (a) and (b) in Fig. \ref{tunnel1}\ respectively
describe the tunneling Hamiltonian for the $\overrightarrow{H}_{T+}$ and $%
\overrightarrow{H}_{T-}$ terms. \ In this diagram a solid line represents a
Cooper pair state in the ground state, whereas a quasiparticle state is
given by a dashed line. \ Only one pair participates in the tunneling
interaction, so only one of the three solid lines is converted to a dashed
line. \ The line of triangles represents the tunneling event and is labeled
with its corresponding $H_{T}$ Hamiltonian, with the direction of the
triangles indicating the direction of the electron tunneling. \ The $%
c_{k}^{\dagger }$ operators, acting on the $L$ or $R$ lead, is rewritten in
terms of the $\gamma $ operators and placed above or below the vertices. \
Since only $\gamma ^{\dagger }$ operators give a nonzero term when acting on
the ground state, the effect of the interaction is to produce final states $%
\Psi _{f}^{L,R}$ with total energy $E_{R}+E_{L}$, and with amplitudes given
at the right of the figure. \ 

\FRAME{ftbpFU}{3.5405in}{3.0727in}{0pt}{\Qcb{First-order Feynmann diagrams
for interaction $\protect\overrightarrow{H}_{T+}$ (a) and $\protect%
\overrightarrow{H}_{T-}$\ (b). \ Solid lines are Cooper-pair states, dashed
lines are quasiparticle excitations, and arrow-lines represents tunneling
interaction. \ Electron operators arising from interaction are displayed
next to vertices. \ }}{\Qlb{tunnel1}}{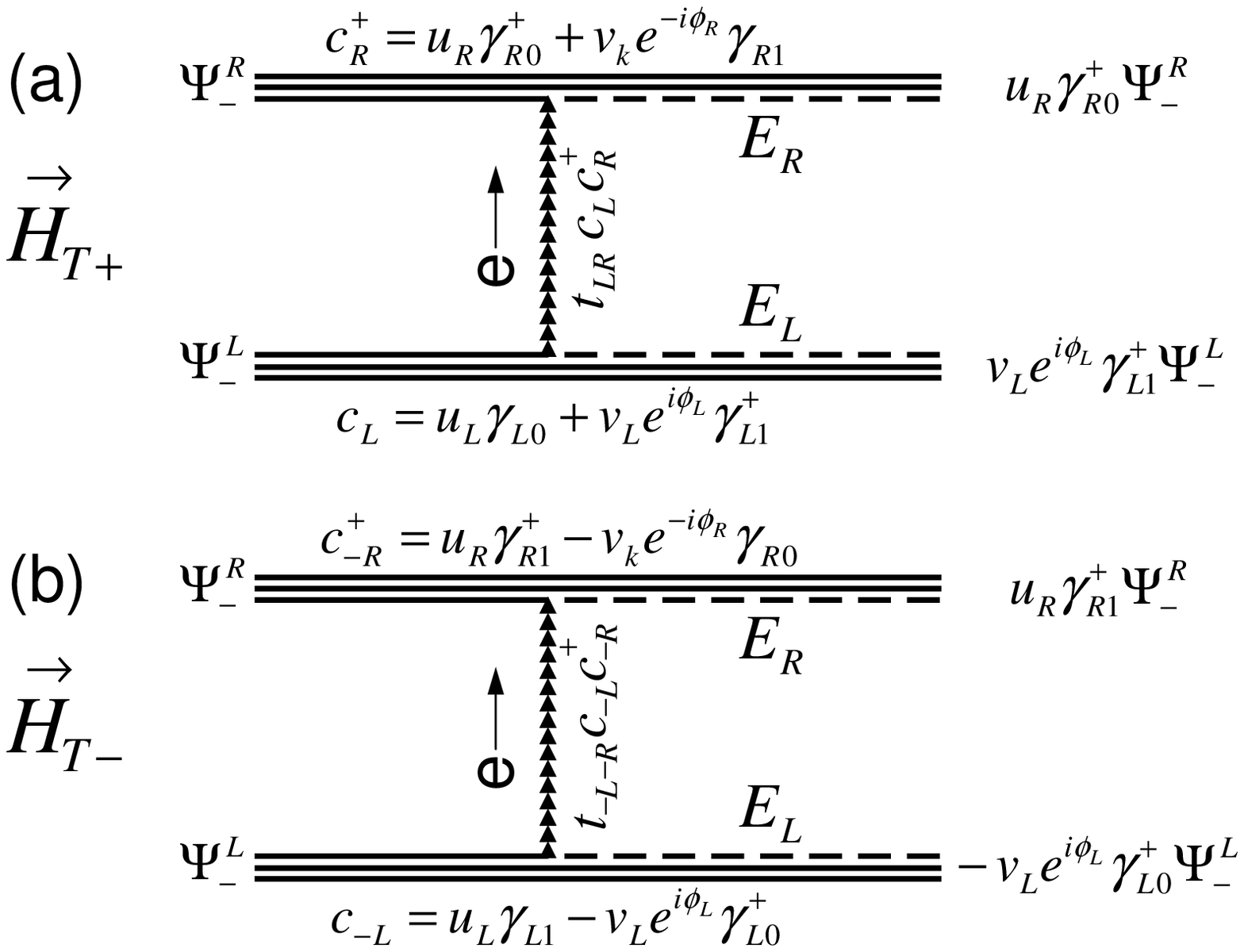}{\special{language
"Scientific Word";type "GRAPHIC";maintain-aspect-ratio TRUE;display
"USEDEF";valid_file "F";width 3.5405in;height 3.0727in;depth
0pt;original-width 5.5564in;original-height 4.817in;cropleft "0";croptop
"1";cropright "1";cropbottom "0";filename 'fig5.eps';file-properties
"XNPEU";}}The two final states in Fig \ref{tunnel1}(a) and (b) are
orthogonal, as well as states involving different values of $L$ and $R$. \
The total current is calculated as an incoherent sum over all possible final
quasiparticle states, under the condition that the total quasiparticle
energy for the final state is equal to the energy gained by the tunneling of
the electron 
\begin{equation}
E_{R}+E_{L}=eV\text{ .}
\end{equation}%
The total current from $L$ to $R$ is given by $e$ multiplied by the
transition rate 
\begin{subequations}
\begin{eqnarray}
\overrightarrow{I_{qp}} &=&e\frac{2\pi }{\hbar }\dsum%
\limits_{L,R}^{(E_{R}+E_{L}=eV)}\left\vert \left\langle \Psi
_{f}^{L}\right\vert \left\langle \Psi _{f}^{R}\right\vert \overrightarrow{H}%
_{T+}+\overrightarrow{H}_{T-}\left\vert \Psi _{-}^{R}\right\rangle
\left\vert \Psi _{-}^{L}\right\rangle \right\vert ^{2} \\
&=&\frac{2\pi e}{\hbar }\dsum\limits_{L,R}^{(E_{R}+E_{L}=eV)}\left[
\left\vert t_{LR}^{{}}\right\vert ^{2}+\left\vert t_{-L-R}^{{}}\right\vert
^{2}\right] (u_{R}^{{}}v_{L}^{{}})^{2} \\
&=&\frac{4\pi e}{\hbar }\left\vert t\right\vert
^{2}N_{0R}N_{0L}\dint\limits_{-\infty }^{\infty }v_{L}^{2}\,d\xi
_{L}\dint\limits_{-\infty }^{\infty }u_{R}^{2}\,d\xi _{R}\text{ }\delta
(eV-E_{L}-E_{R})\text{ ,}  \label{Iqp3}
\end{eqnarray}%
where in the last equation we have expressed the conservation of energy with
a Dirac $\delta $-function, and have assumed matrix elements $\left\vert
t\right\vert ^{2}$ of constant strength. \ Because $E(\xi _{k})=E(-\xi _{k})$
and $u_{k}(\xi _{k})=v_{k}(-\xi _{k})$, one finds 
\end{subequations}
\begin{subequations}
\begin{eqnarray}
\overrightarrow{I_{qp}} &=&\frac{4\pi e}{\hbar }\left\vert t\right\vert
^{2}N_{0R}N_{0L}\dint\limits_{0}^{\infty }(v_{L}^{2}\,+u_{L}^{2})d\xi
_{L}\dint\limits_{0}^{\infty }(u_{R}^{2}+v_{R}^{2})\,d\xi _{R}\text{ }\delta
(eV-E_{L}-E_{R})  \notag \\
&=&\frac{4\pi e}{\hbar }\left\vert t\right\vert
^{2}N_{0R}N_{0L}\dint\limits_{0}^{\infty }d\xi _{L}\dint\limits_{0}^{\infty
}d\xi _{R}\text{ }\delta (eV-E_{L}-E_{R})\text{ .}  \label{Iqp5}
\end{eqnarray}

This result is equivalent to the standard \textquotedblleft semiconductor
model\textquotedblright\ of the quasiparticle current, which predicts no
current for $V<2\Delta /e$, a rapid rise of current at $2\Delta /e$, and
then a current proportional to $V$ at large voltages. \ Note that Eq. \ref%
{Iqp3} has a sum over the occupation probability $v_{L}^{2}$ of the pair
state and the occupation probability $u_{R}^{2}$ of a hole-pair state, as is
expected given the operators $c_{L}^{{}}c_{R}^{\dag }$ in the tunneling
Hamiltonian. \ The final result of Eq. \ref{Iqp5} does not have these
factors because the occupation probability is unity when summed over the $%
\pm \xi _{k}$ states. \ \ 

It is convenient to express the tunneling matrix element in terms of the
normal-state resistance of the junction, obtained by setting $\Delta =0,$
with the equation 
\end{subequations}
\begin{subequations}
\begin{eqnarray}
1/R_{N} &\equiv &\overrightarrow{I_{qp}}/V \\
&=&\frac{4\pi e}{\hbar }\left\vert t\right\vert
^{2}N_{0R}N_{0L}\dint\limits_{0}^{\infty }d\xi _{L}\dint\limits_{0}^{\infty
}d\xi _{R}\text{ }\delta (eV-\xi _{L}-\xi _{R})/V \\
&=&\frac{4\pi e^{2}}{\hbar }\left\vert t\right\vert ^{2}N_{0R}N_{0L}\text{ .}
\end{eqnarray}

We now calculate the tunneling current with second-order perturbation
theory. \ The tunneling Hamiltonian, taken to second order, gives 
\end{subequations}
\begin{equation}
H_{T}^{(2)}=\tsum_{i}H_{T}\frac{1}{\epsilon _{i}}H_{T}\text{ ,}
\end{equation}%
where $\epsilon _{i}$ is the energy of the intermediate state $i$. \ Because
the terms in $H_{T}$ have\ both $\gamma ^{\dagger }$ and $\gamma $
operators, the second-order Hamiltonian gives a nonzero expectation value
for the ground state. \ This is unlike the first-order theory, which
produces current only through the real creation of quasiparticles. \ 

\FRAME{ftbpFU}{3.5405in}{2.514in}{0pt}{\Qcb{Second-order Feynman diagrams
for the transfer of two electrons across the junction. \ Only nonzero
operators are displayed next to vertices.}}{\Qlb{tunnel2}}{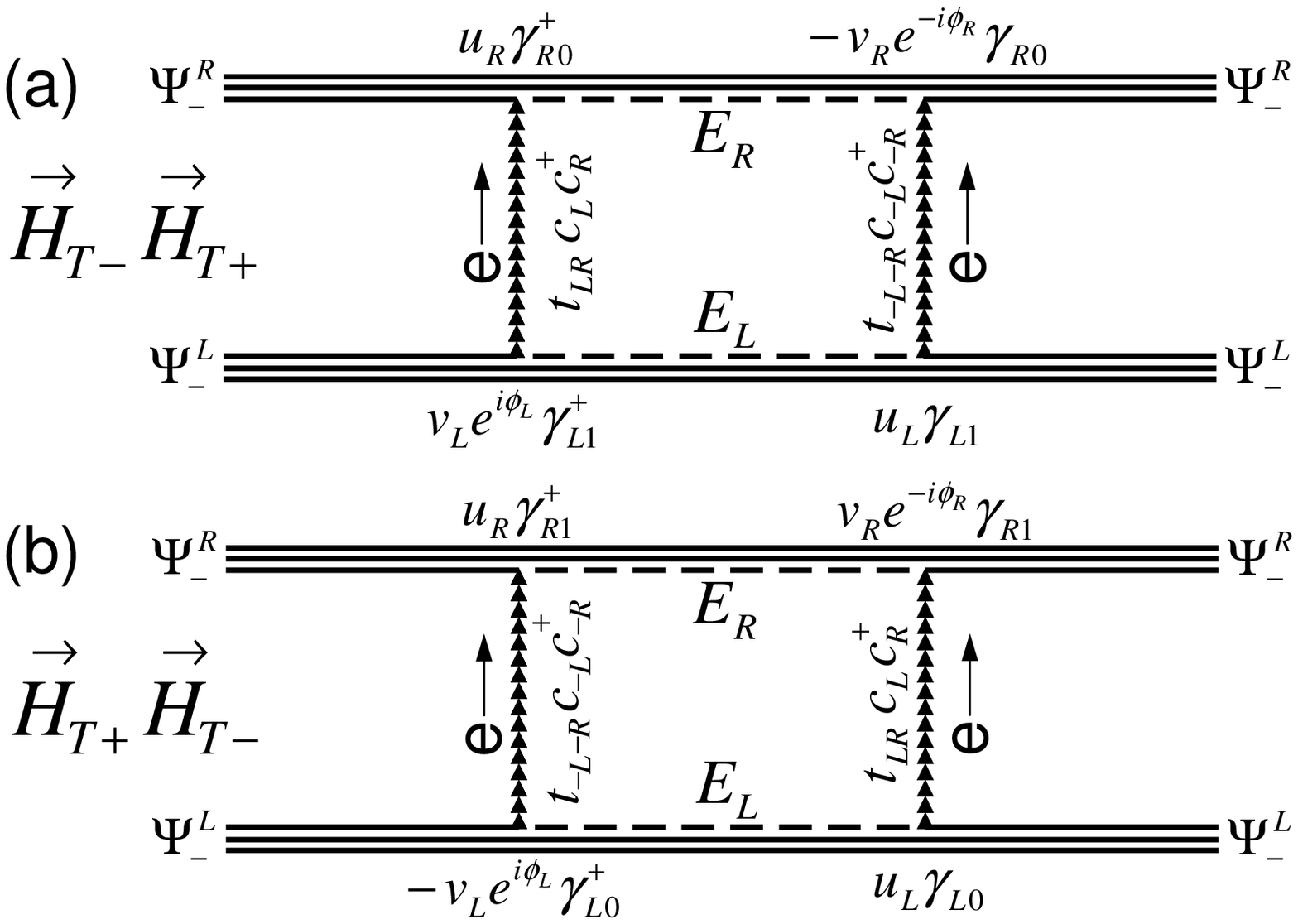}{\special%
{language "Scientific Word";type "GRAPHIC";maintain-aspect-ratio
TRUE;display "USEDEF";valid_file "F";width 3.5405in;height 2.514in;depth
0pt;original-width 6.6245in;original-height 4.6864in;cropleft "0";croptop
"1";cropright "1";cropbottom "0";filename 'fig6.eps';file-properties
"XNPEU";}}

Because $H_{T}$ has terms that transfer charge in both directions, $%
H_{T}H_{T}$ will produce terms which transfer two electrons to the right,
two to the left, and with no net transfer. \ \ With no transfer, a
calculation of the second-order energy gives a constant value, which has no
physical effect. \ We first calculate terms for transfer to the right from $(%
\overrightarrow{H}_{T+}+\overrightarrow{H}_{T-})(\overrightarrow{H}_{T+}+%
\overrightarrow{H}_{T-})$, which gives nonzero expectation values only for $%
\overrightarrow{H}_{T+}\overrightarrow{H}_{T-}+\overrightarrow{H}_{T-}%
\overrightarrow{H}_{T+}$. \ The Feynman diagrams for these two terms are
given in Fig. \ref{tunnel2}(a) and (b), where we have displayed only the
amplitudes from the nonzero operators. \ The expectation value of these two
Hamiltonian terms is given by 
\begin{subequations}
\begin{eqnarray}
\left\langle \overrightarrow{H_{T}^{(2)}}\right\rangle
&=&-\dsum\limits_{L,R}\left\langle \Psi _{-}^{R}\right\vert \left\langle
\Psi _{-}^{L}\right\vert (v_{R}^{{}}e^{-i\phi _{R}}u_{L}^{{}})  \notag \\
&&\qquad \times \frac{\gamma _{R0}^{{}}\gamma _{L1}^{{}}\gamma
_{R0}^{\dagger }\gamma _{L1}^{\dagger }t_{LR}^{{}}t_{-L-R}^{{}}+\gamma
_{R1}^{{}}\gamma _{L0}^{{}}\gamma _{R1}^{\dagger }\gamma _{L0}^{\dagger
}t_{-L-R}^{{}}t_{LR}^{{}}}{E_{R}+E_{L}}  \notag \\
&&\qquad \times (u_{R}^{{}}v_{L}^{{}}e^{i\phi _{L}})\left\vert \Psi
_{-}^{L}\right\rangle \left\vert \Psi _{-}^{R}\right\rangle \\
&=&-2\left\vert t\right\vert ^{2}e^{i(\phi _{L}-\phi
_{R})}\dsum\limits_{L,R}(v_{R}^{{}}u_{R}^{{}})(u_{L}^{{}}v_{L}^{{}})\frac{1}{%
E_{R}+E_{L}} \\
&=&-2\left\vert t\right\vert ^{2}e^{i\delta
}N_{0R}N_{0L}\dint\limits_{-\infty }^{\infty }d\xi _{R}\dint\limits_{-\infty
}^{\infty }d\xi _{L}\frac{\Delta }{E_{R}}\frac{\Delta }{E_{L}}\frac{1}{%
E_{R}+E_{L}} \\
&=&-\frac{\hbar \Delta }{2\pi e^{2}R_{N}}e^{i\delta }\dint\limits_{-\infty
}^{\infty }d\theta _{R}\dint\limits_{-\infty }^{\infty }d\theta _{L}\frac{1}{%
\cosh \theta _{R}+\cosh \theta _{L}} \\
&=&-\frac{\hbar \Delta }{2\pi e^{2}R_{N}}e^{i\delta }\left( \frac{\pi }{2}%
\right) ^{2}\text{ ,}  \label{HT2}
\end{eqnarray}%
where we have used $t_{LR}^{\ast }=t_{-L-R}^{{}}$ and assumed the same gap $%
\Delta $ for both superconductors. \ \ A similar calculation for transfer to
the left gives the complex conjugate of Eq. \ref{HT2}. \ The sum of these
two energies gives the Josephson energy $U_{J}$, and using Eq. \ref{UtoI},
the Josephson current $I_{J}$, 
\end{subequations}
\begin{eqnarray}
U_{J} &=&-\frac{1}{8}\frac{R_{K}}{R_{N}}\Delta \cos \delta \text{ ,}
\label{EqUjAB} \\
I_{J} &=&\frac{\pi }{2}\frac{\Delta }{eR_{N}}\sin \delta \text{ ,}
\label{IJAB}
\end{eqnarray}%
where $R_{K}=h/e^{2}$ is the resistance quantum. \ Equation \ref{IJAB} is
the standard Ambegaokar-Baratoff formula\cite{AB} for the Josephson current
at zero temperature. \ 

The Josephson current is a dissipationless current because it arises from a
new ground state of the two superconductors produced by the tunneling
interaction. \ This behavior is in contrast with quasiparticle tunneling,
which is dissipative because it produces excitations. \ It is perhaps
surprising that a new ground state can produce charge transfer through the
junction. \ This is possible only because the virtual quasiparticle
excitations are both electrons and holes: the electron-part tunnels first
through the junction, then the hole-part tunnels back. \ Only states of
energy $\Delta $ around the Fermi energy are both electron- and hole-like,
as weighted by the $(v_{R}^{{}}u_{R}^{{}})(u_{L}^{{}}v_{L}^{{}})$ term in
the integral. \ 

The form of the Josephson Hamiltonian can be understood readily by noting
that the second-order Hamiltonian, 
\begin{eqnarray}
\overrightarrow{H}_{T+}\overrightarrow{H}_{T-} &\sim &\left\vert
t\right\vert ^{2}\dsum\limits_{L,R}c_{L}^{{}}c_{-L}^{{}}c_{R}^{\dag
}c_{-R}^{\dag } \\
&=&\frac{\left\vert t\right\vert ^{2}}{2}\dsum\limits_{L,R}(\sigma
_{xL}\sigma _{xR}+\sigma _{yL}\sigma _{yR})\text{ ,}
\end{eqnarray}%
corresponds to the pair-scattering Hamiltonian of Eq. \ref{Hpair}. \
Comparing with the gap-equation solution, one expects $U_{J}\sim \left\vert
t\right\vert ^{2}\Delta \cos \delta $, where the $\cos \delta $ term arises
from the spin-spin interaction in the x-y plane.

We would like to make a final comment on a similarity between the BCS theory
and the Josephson effect. \ In both of these derivations we see that a
dissipative process that is described in first-order perturbation theory,
such as phonon scattering or quasiparticle tunneling, produces in second
order a new collective superfluid behavior. \ This collective behavior
emerges from a virtual excitation of the dissipative process. \ Dissipation
is normally considered undesirable, but by designing systems to \textit{%
maximize} dissipation, it may be possible to discover new quantum collective
behavior.

With this understanding of the Josephson effect and quasiparticle tunneling,
how accurate is the description of the Josephson junction with the
Hamiltonian corresponding to Eq. \ref{EqUjAB}? There are several issues that
need to be considered. \ 

First, quasiparticle tunneling is a dissipative mechanism that produces
decoherence. \ Although it is predicted to be absent for $V<2\Delta /e$,
measurements of real junctions show a small subgap current. \ This current
is understood to arise from multiple Andreev reflections, which are
described as higher-order tunneling processes. \ We thus need a description
of the tunnel junction that easily predicts these processes for arbitrary
tunneling matrix elements. \ This is especially needed as real tunnel
junctions do not have constant matrix elements, as assumed above. \
Additionally, we would like to know whether a small number of major
imperfections, such as \textquotedblleft pinhole\textquotedblright\ defects,
will strongly degrade the coherence of the qubit.

Second, quasiparticle tunneling has been predicted for an arbitrary DC
voltage across the junction. \ However, the qubit state has $\langle
V\rangle =0$, but may excite quasiparticles with AC voltage fluctuations. \
This situation is difficult to calculate with perturbation theory. \ \ \ In
addition, is it valid to estimate decoherence from quasiparticles at zero
voltage simply from the junction resistance at subgap voltages?

Third, how will the Josephson effect and the qubit Hamiltonian be modified
under this more realistic description of the tunnel junction? \ 

All of these questions and difficulties arise because perturbation theory
has been used to describe the ground state of the Josephson junction. \ The
BCS theory gives basis states that best describe quasiparticle tunneling for
large voltages, not for $V\rightarrow 0$. \ \ A theory is needed that solves
for the Josephson effect \textit{exactly}, with this solution then providing
the basis states for understanding quasiparticle tunneling around $V=0$. \
This goal is fulfilled by the theory of quasiparticle bound states, which we
will describe next. \ 

\section{\label{Andreev}The Josephson Effect, Derived from Quasiparticle
Bound States}

We begin our derivation of an exact solution for the Josephson effect with
an extremely powerful idea from mesoscopic physics: electrical transport can
be calculated under very general conditions by summing the current from a
number of \textit{independent} \textquotedblleft conduction
channels\textquotedblright , with the transport physics of each conduction
channel determined only by its channel transmission probability $\tau _{i}$%
\cite{Beenakker91,Beenakker97}. \ For a Josephson junction, the total
junction current $I_{j}$ can be written as a sum over all channels $i$ 
\begin{equation}
I_{j}=\tsum_{i}I_{j}(\tau _{i})\text{ ,}
\end{equation}%
where $I_{j}(\tau )$ is the current for a single channel of transmission $%
\tau $, which may be solved for theoretically. \ For a tunnel junction, the
number of channels is estimated as the junction area divided by the channel
area $(\lambda _{f}/2)^{2}$, where $\lambda _{f}$ is the Fermi wavelength of
the electrons. \ Of course, the difficulty of determining the distribution
of the channel transmissions still remains. \ This often may be estimated
from transport properties, and under some situations can be predicted from
theory\cite{Naveh,Dorokov,Schep}. \ 

Because transport physics is determined \textit{only} by scattering
parameterized by $\tau $, we may make two simplifying assumptions: the
transport can be solved for using plane waves, and the scattering from the
tunnel junction can be described by a delta function. \ The general theory
has thus been transformed into the problem of one-dimensional scattering
from a delta function, and an exact solution can be found by using a simple
and clear physical picture. \ \ 

\FRAME{ftbpFU}{4.0413in}{1.0188in}{0pt}{\Qcb{Plot of potential \textit{vs.}
coordinate $x$ with a positive delta-function tunnel barrier $V_{0}\protect%
\delta (x)$. \ Scattering of plane wave states is shown in (a), whereas (b)
is a plot of the bound-state wavefunction. \ The delta-function barrier is
negative in (b), as required for producing a bound state. \ }}{\Qlb{%
figscatter}}{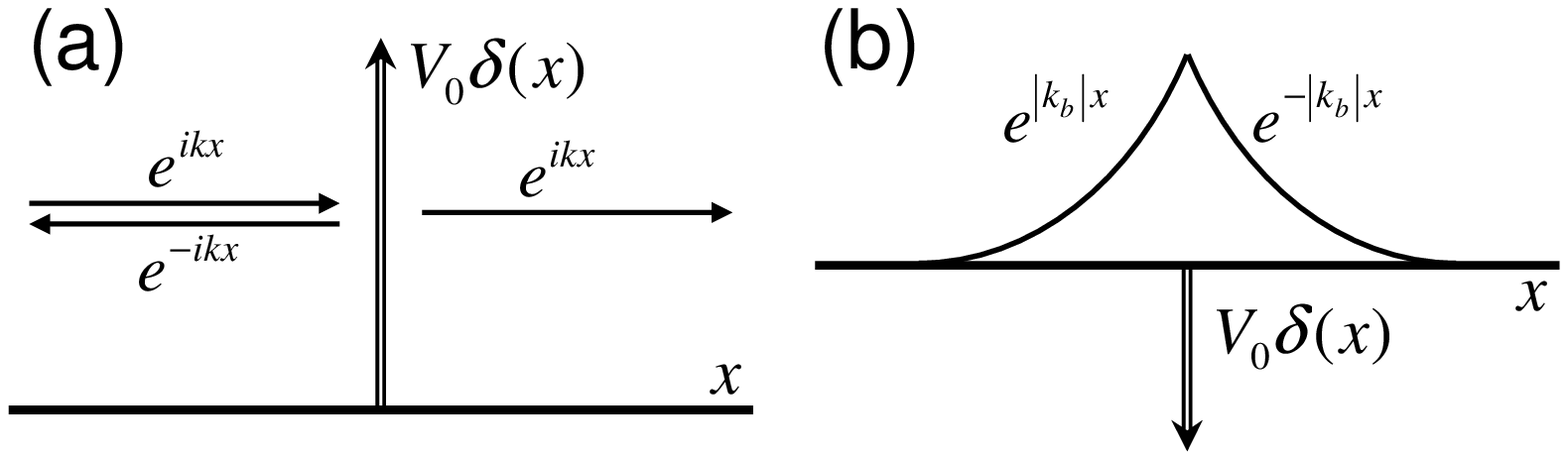}{\special{language "Scientific Word";type
"GRAPHIC";maintain-aspect-ratio TRUE;display "USEDEF";valid_file "F";width
4.0413in;height 1.0188in;depth 0pt;original-width 6.2906in;original-height
1.5541in;cropleft "0";croptop "1";cropright "1";cropbottom "0";filename
'fig7.eps';file-properties "XNPEU";}}

Central to understanding the Josephson effect will be the quasiparticle
bound state. \ To understand how to calculate a bound state\cite%
{Furusaki,Kuplev}, we will first consider a normal-metal tunnel junction and
with a $\delta $-function barrier $V_{0}\overline{\delta }(x)$, as
illustrated in Fig. \ref{figscatter}. \ For an electron of mass $m$ and
wavevector $k$, the wavefunctions on the left and right side of the barrier
are 
\begin{eqnarray}
\Psi _{L} &=&Ae^{ikx}+Be^{-ikx}\text{ ,}  \label{NLwave} \\
\Psi _{R} &=&Ce^{ikx}\text{ ,}  \label{NRwave}
\end{eqnarray}%
where $A$, $B$, and $C$ are respectively the incident, reflected, and
transmitted electron amplitudes. \ From the continuity equations%
\begin{eqnarray}
\Psi _{L} &=&\Psi _{R}  \label{Cont1} \\
\frac{d\Psi _{L}}{dx} &=&\frac{d\Psi _{R}}{dx}-\frac{2mV_{0}}{\hbar ^{2}}%
\Psi _{R}  \label{Cont2}
\end{eqnarray}%
evaluated at $x=0$, the amplitudes are related by%
\begin{eqnarray}
A+B &=&C  \label{Psi} \\
ikA-ikB &=&ikC-\frac{2mV_{0}}{\hbar ^{2}}C\text{ .}  \label{dPsi}
\end{eqnarray}%
The transmission amplitude and the probability are 
\begin{eqnarray}
\frac{C}{A} &=&\frac{1}{1+i\eta }\text{ ,} \\
\tau &=&\left\vert \frac{C}{A}\right\vert ^{2}=\frac{1}{1+\eta ^{2}}\text{ ,}
\end{eqnarray}%
where $\eta =mV_{0}/\hbar ^{2}k$. \ The bound state can be determined by
finding the pole in the transmission amplitude. \ A pole describes how a
state of finite amplitude may be formed around the scattering site with zero
amplitude of the incident wavefunction, which is the definition of a bound
state. \ The pole at $\eta =i$ gives $k_{b}=-imV_{0}/\hbar ^{2}$, and a
wavefunction around the scattering site $\Psi _{R}=Ce^{(mV_{0}/\hbar ^{2})x}$%
. \ This describes a bound state only when $V_{0}$ is negative, as expected.

\FRAME{ftbpFU}{3.5405in}{1.6579in}{0pt}{\Qcb{Plot of quasiparticle energies $%
E_{\protect\kappa }$ verses momentum $\protect\kappa $ near the $\pm k_{f}$
Fermi surfaces. \ The two-component\ eigenfunctions are also displayed for
each of the four energy bands. \ \ Also indicated are the quasiparticle
states A-E used for the bound-state calculation. }}{\Qlb{figsBdG}}{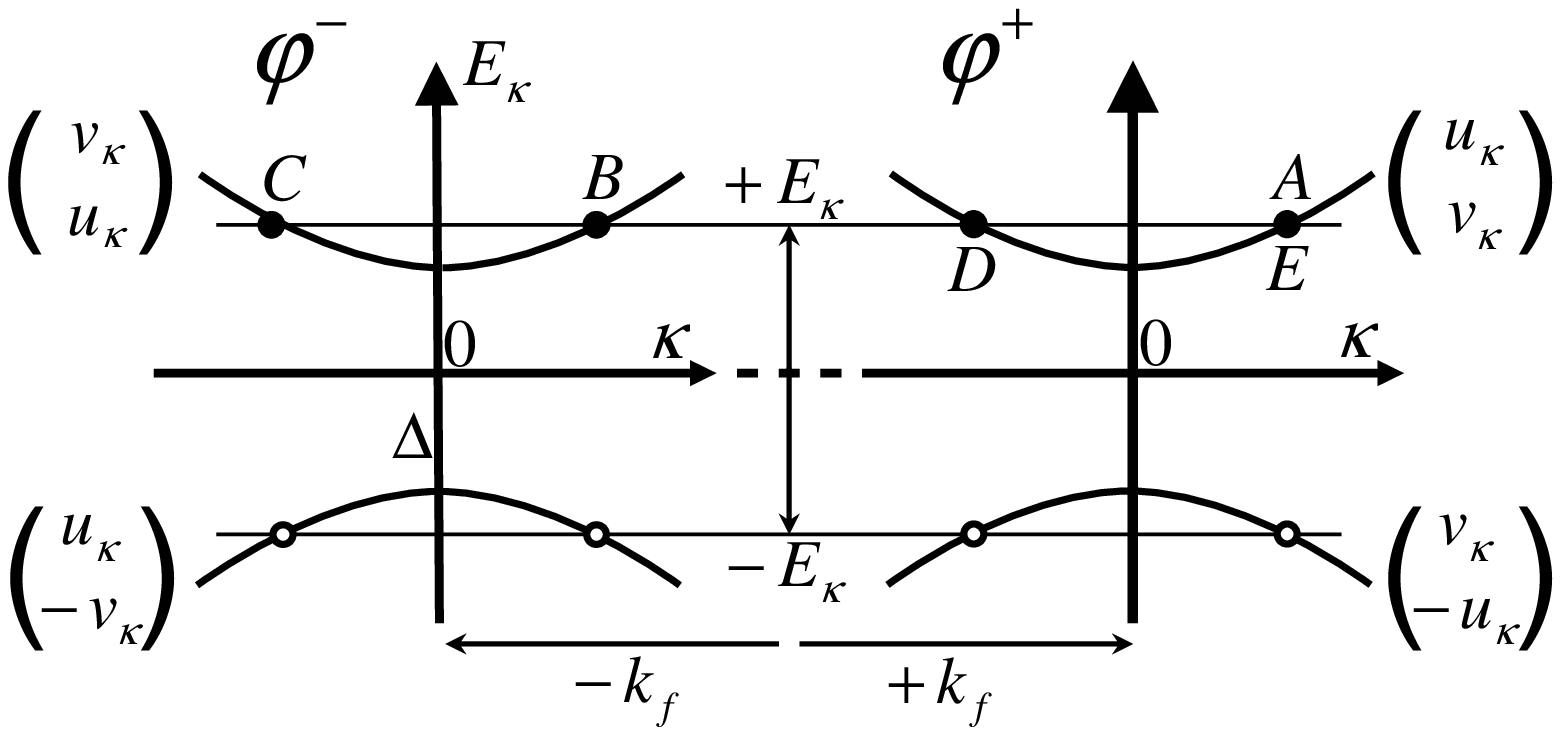}{%
\special{language "Scientific Word";type "GRAPHIC";maintain-aspect-ratio
TRUE;display "USEDEF";valid_file "F";width 3.5405in;height 1.6579in;depth
0pt;original-width 6.1981in;original-height 2.687in;cropleft "0";croptop
"1";cropright "1";cropbottom "0";filename 'fig8.eps';file-properties
"XNPEU";}}

A superconducting tunnel junction will also have bound states of
quasiparticles excitations. \ These bound states describe the Josephson
effect since virtual quasiparticle tunneling was necessary for the
perturbation calculation in the last section. \ The Bogoliubov-deGennes
equations describe the spatial wavefunctions, whose eigenstates are given by
the solution of the Hamiltonian 
\begin{equation}
H\varphi ^{\pm }e^{i\kappa x}=\left( \pm k_{f}\frac{\hbar ^{2}\kappa }{m}%
\sigma _{z}+\Delta \sigma _{x}\right) \varphi ^{\pm }e^{i\kappa x}\text{ ,}
\end{equation}%
where $\varphi ^{\pm }e^{i\kappa x}$ are the slowly varying spatial
amplitudes of the exact wavefunction $\varphi ^{\pm }e^{i\kappa x}e^{\pm
ik_{f}x}$. \ As illustrated in Fig. \ref{figsBdG}, the $\pm k_{f}\kappa $
term corresponds to the kinetic energy at the $\pm k_{f}$ Fermi surfaces
using the approximation $(k_{f}+\kappa )^{2}/2\simeq \mathrm{const.}%
+k_{f}\kappa $ . \ As expected for a spin-type Hamiltonian, the two
eigenvalues are 
\begin{equation}
E_{\kappa }=\pm (\xi _{\kappa }^{2}+\Delta ^{2})^{1/2}\text{ ,}
\end{equation}%
where $\xi _{\kappa }=\hbar ^{2}k_{f}\kappa /m$ is the kinetic energy of the
quasiparticle referred to the Fermi energy. \ The eigenvectors are also
displayed in Fig. \ref{figsBdG}, where $u_{\kappa }$ and $v_{\kappa }$ are
given by Eqs. \ref{equ} and \ref{eqv}. Because the two energy bands
represent quasiparticle excitations, the lower band is normally filled and
its excitations correspond to the creation of hole states. \ \ \ \ \ 

We can solve for the quasiparticle bound states by first writing down the
scattering wavefunctions in the left and right superconducting electrodes. \
An incoming quasiparticle state, point A in Fig. \ref{figsBdG}, is reflected
off the tunnel barrier to states B and C and is transmitted to states D and
E \cite{states}. \ The wavefunctions are then given by 
\begin{eqnarray}
\Psi _{L} &=&A\binom{u}{ve^{i\phi _{L}}}e^{i\kappa x}+B\binom{v}{ue^{i\phi
_{L}}}e^{i\kappa x}+C\binom{u}{ve^{i\phi _{L}}}e^{-i\kappa x} \\
\Psi _{R} &=&D\binom{v}{ue^{i\phi _{R}}}e^{-i\kappa x}+E\binom{u}{ve^{i\phi
_{R}}}e^{i\kappa x}\text{ ,}
\end{eqnarray}%
where we have used the relations $v\equiv v_{\kappa }=u_{-\kappa }$ and $%
u\equiv u_{\kappa }=v_{-\kappa }$, and we have included the phases $\phi
_{L} $ and $\phi _{R}$ of the two states. \ \ The continuity conditions Eqs. %
\ref{Cont1} and \ref{Cont2}, solved for both the components of the spin
wavefunction, gives the matrix equation

\begin{equation}
A%
\begin{pmatrix}
u \\ 
v \\ 
u \\ 
v%
\end{pmatrix}%
=%
\begin{pmatrix}
-v & -u & v & u \\ 
-u & -v & ue^{-i\delta } & ve^{-i\delta } \\ 
-v & u & -v(1-i2\eta ) & u(1+i2\eta ) \\ 
-u & v & -ue^{-i\delta }(1-i2\eta ) & ve^{-i\delta }(1+i2\eta )%
\end{pmatrix}%
\begin{pmatrix}
B \\ 
C \\ 
D \\ 
E%
\end{pmatrix}%
\text{ .}
\end{equation}%
The scattering amplitudes for $B$-$E$ have poles given by the solution of 
\begin{equation}
(u^{4}+v^{4})(1+\eta ^{2})-2(uv)^{2}(\eta ^{2}+\cos \delta )=0\text{ .}
\end{equation}%
Using the relations $u^{2}+v^{2}=1$, $E_{J}=E_{k}=\Delta /2uv$, and $\tau
=1/(1+\eta ^{2})$, the energies of the quasiparticle bound states are%
\begin{equation}
E_{J\pm }=\pm \Delta \lbrack 1-\tau \sin ^{2}(\delta /2)]^{1/2}\text{ .}
\label{EqE}
\end{equation}%
Because these two states have energies less than the gap energy $\Delta $,
they are energetically \textquotedblleft bound\textquotedblright\ to the
junction and thus have wavefunctions that are localized around the junction.

\FRAME{ftbpFU}{2.5374in}{2.0167in}{0pt}{\Qcb{Plot of quasiparticle
bound-state energies $E_{J-}$ and $E_{J+}$ vs. the phase difference $\protect%
\delta $ across the junction, for three values of tunneling transmission $%
\protect\tau $. \ Quasiparticles are produced by vertical transitions from
the $E_{J-}$\ to $E_{J+}$\ band. As indicated by the arrow, the energy gap $%
E_{J+}-E_{J-}$ is always greater than $\protect\sqrt{2}\Delta $ at $\protect%
\delta =\protect\pi /2$.}}{\Qlb{figsBS}}{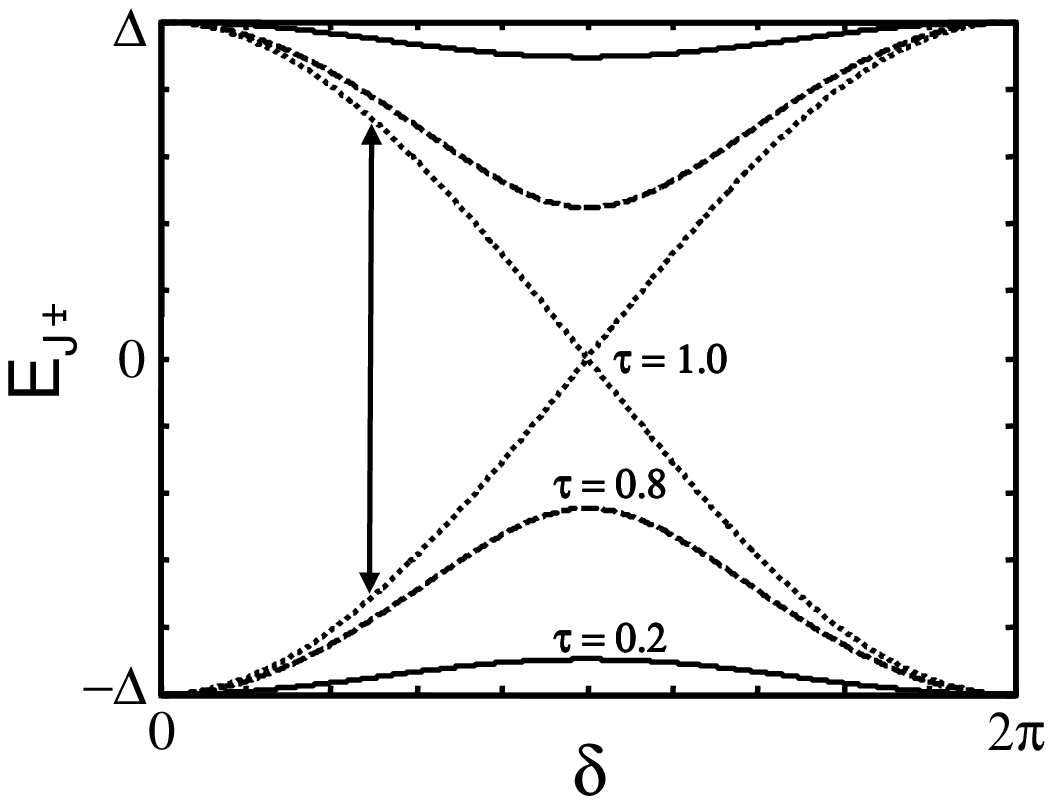}{\special{language
"Scientific Word";type "GRAPHIC";maintain-aspect-ratio TRUE;display
"USEDEF";valid_file "F";width 2.5374in;height 2.0167in;depth
0pt;original-width 4.0932in;original-height 3.243in;cropleft "0";croptop
"1";cropright "1";cropbottom "0";filename 'fig9.eps';file-properties
"XNPEU";}}

The dependence of the quasiparticle bound-state energies on junction phase
is plotted in Fig. \ref{figsBS} for several values of $\tau $. \ The ground
state$\ $is normally filled, similar to the filling of quasiparticle states
of negative energy. \ The energy $E_{J-}$ corresponds to the Josephson
energy, as can be checked in the limit $\tau \rightarrow 0$ to give%
\begin{equation}
E_{J-}\simeq -\Delta +\frac{\Delta \tau }{4}-\frac{\Delta \tau }{4}\cos
\delta \text{ .}
\end{equation}%
This result is equivalent to Eq. \ref{EqUjAB} after noting that the
normal-state conductance of a single channel is $1/R_{N}=2\tau /R_{K}$. \ 

The current of each bound state is given by the derivative of its energy 
\begin{equation}
I_{J\pm }=\frac{2\pi }{\Phi _{0}}\frac{\partial E_{J\pm }}{\partial \delta }%
\text{ ,}  \label{EqIj}
\end{equation}%
in accord with Eq. \ref{UtoI}. \ Since the curvature of the upper band is
opposite to that of the lower band, the currents of the two bands have
opposite sign $I_{J+}=-I_{J-}$. \ For level populations of the two states
given by $f_{\pm }$, the average Josephson current is $\left\langle
I_{J}\right\rangle =I_{J-}(f_{-}-f_{+})$. \ For a thermal population, $%
f_{\pm }$ are given by Fermi distributions, and the Josephson current in the
tunnel junction limit gives the expected Ambegaokar-Baratoff result 
\begin{eqnarray}
\left\langle I_{J}\right\rangle &=&\frac{\pi }{2}\frac{\Delta (T)}{eR_{N}}%
\sin \delta \cdot \left( \frac{1}{e^{-\Delta /kT}+1}-\frac{1}{e^{\Delta
/kT}+1}\right) \\
&=&\frac{\pi }{2}\frac{\Delta (T)}{eR_{N}}\tanh (\Delta /2kT)\sin \delta 
\text{ .}
\end{eqnarray}%
\ 

\section{\label{QPzener}Generation of Quasiparticles from Nonadiabatic
Transitions}

In this description of the Josephson junction, the Josephson effect arises
from a quasiparticle bound state at the junction. \ Two bound states exist
and have energies $E_{J+}$ and $E_{J-}$, with the Josephson current from the
excited state being of opposite sign from that of the ground state. \ We
will discuss here the small-voltage limit\cite{Averin,Bratus}, which can be
fully understood within a semiclassical picture by considering that a linear
increase in $\delta $ produces nonadiabatic transitions between the two
states. \ 

The junction creates \textquotedblleft free quasiparticles\textquotedblright
, those with $E\geq \Delta $, via a two-step process. \ First, a transition
is made from the ground to the excited bound state. \ This typically occurs
because a voltage is placed across the junction, and the linear change of $%
\delta $ causes the ground state not to adiabatically stay in that state. \
For a high-transmission channel, the transition is usually made around $%
\delta \approx \pi $, where the energy difference between the states is the
lowest and the band bending is the highest. \ Because this excited state
initially has energy less than $\Delta $, the state remains bound until the
phase changes to $2\pi $ and the energy of the quasiparticle is large enough
to become unbound and diffuse away from the junction. \ The quasiparticle
generation rate is thus governed by $d\delta /dt$\ and will increase as $V$
increases. \ \ 

\FRAME{ftbpFU}{4.2436in}{1.3898in}{0pt}{\Qcb{Plot of semiclassical solutions
for the tunneling through a barrier (a) and tunneling through an energy gap
(b). \ Imaginary solutions to $k$ and $\protect\delta $ are used to
calculate the tunneling rates. \ }}{\Qlb{figWKB}}{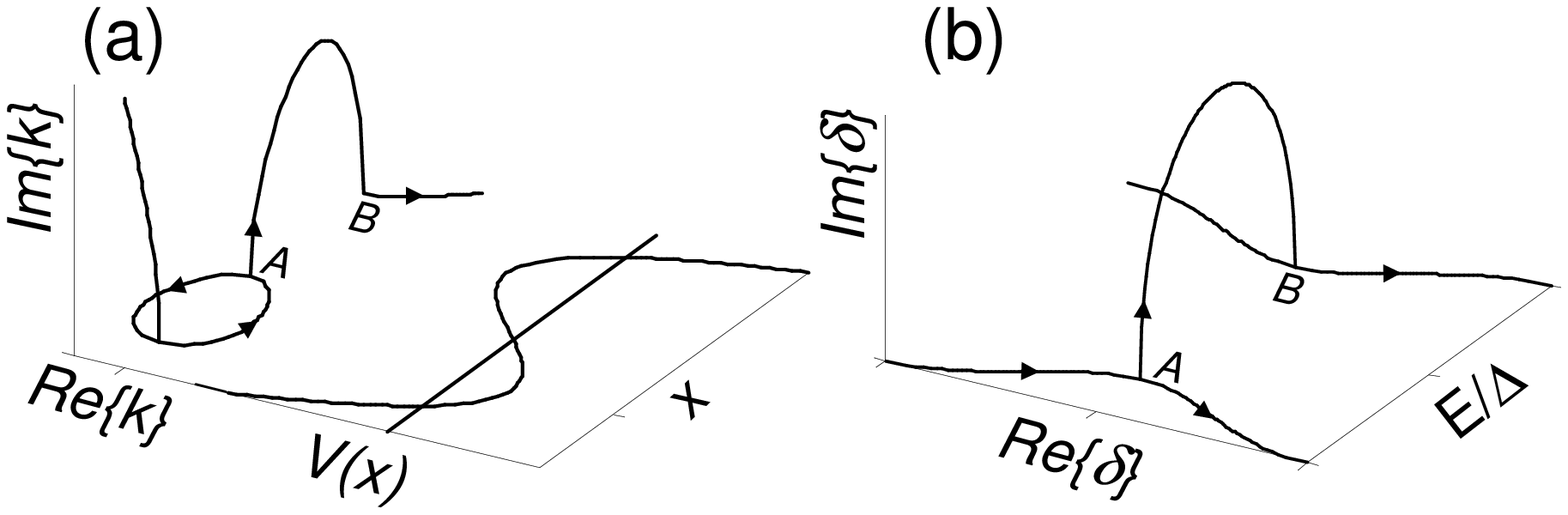}{\special%
{language "Scientific Word";type "GRAPHIC";maintain-aspect-ratio
TRUE;display "USEDEF";valid_file "F";width 4.2436in;height 1.3898in;depth
0pt;original-width 7.2627in;original-height 2.348in;cropleft "0";croptop
"1";cropright "1";cropbottom "0";filename 'fig10.eps';file-properties
"XNPEU";}}

The quasiparticle transition rate can be predicted using a simple
semi-classical method. \ We will first review WKB tunneling in order to
later generalize this calculation to energy tunnelling. \ In Fig. \ref%
{figWKB}(a), we plot a cubic potential $V(x)$ versus $x$ and its solution $%
k^{2}=2m[E-V(x)]/\hbar ^{2}$. \ The solution for $k$ is real or imaginary
depending on whether $E$ is greater or less than $V(x)$. \ A semi-classical
description of the system is the particle oscillating in the well, as
described by the loop in the solution of $\mathrm{Re}\{k\}$. \ A solution in
the imaginary part of $k$ connects a turning point on this loop, labeled A,
with the turning point of the free-running solution, labeled B. \ The
probability of tunneling each time the trajectory passes point A is given by
the standard WKB integral of the imaginary action 
\begin{eqnarray}
W &=&\exp [-2S]  \label{EqW} \\
S &=&(1/\hbar )\left\vert \int dx\func{Im}p\right\vert  \label{Eqp} \\
&=&\left\vert \int_{x_{A}}^{x_{B}}dx\func{Im}k\text{ }\right\vert \text{.}
\label{Eqk}
\end{eqnarray}

The transition rate for a nonadiabatic change in a state may be calculated
in a similar fashion. \ In Fig. \ref{figWKB}(b) we plot the solution of Eq. %
\ref{EqE} for $\delta $ versus $E$. \ In the \textquotedblleft
forbidden\textquotedblright\ region of energy $\left\vert E\right\vert
<\Delta \sqrt{1-\tau }$, the solution of $\delta $ has an imaginary
component. \ As the bias of the system changes and the system trajectory
moves past point A, then this state can tunnel to point B via the connecting
path in the imaginary part of $\delta $. \ The probability for this event is
given by Eq. \ref{EqW} with $S$ given by the integral of the imaginary action%
\begin{equation}
S=(1/\hbar )\int dE\,t_{\mathrm{imag}}\text{ ,}
\end{equation}%
where we define an imaginary time by%
\begin{equation}
t_{\mathrm{imag}}=\left\vert \frac{\func{Im}\delta }{d\delta /dt}\right\vert 
\text{ .}
\end{equation}%
Rewriting Eq. \ref{EqE} as $\left( E/\Delta \right) ^{2}=1-\tau (1-\cos
\delta )/2$ and using $d\delta /dt=(2e/\hbar )V$, one finds the action is
given by the integral%
\begin{eqnarray}
S &=&\frac{\Delta }{2eV}\dint\nolimits_{-\sqrt{1-\tau }}^{\sqrt{1-\tau }%
}d\epsilon \,\func{Im}\left\{ -\arccos \left[ 1+(\varepsilon ^{2}-1)2/\tau %
\right] \right\}  \label{EqSE} \\
&\simeq &\frac{\Delta }{eV}\times \left\{ 
\begin{array}{l}
(1-\tau )\pi /2 \\ 
\ln (2/\tau ) \\ 
(1-\tau )\left[ \ln (2/\tau )+\sqrt{\tau }(\pi /2-\ln 2)\right] \text{ ,}%
\end{array}%
\right. \left. 
\begin{array}{c}
(\tau \rightarrow 1) \\ 
(\tau \rightarrow 0) \\ 
(\mathrm{interp.})%
\end{array}%
\right.  \notag
\end{eqnarray}%
where the last interpolation formula approximates well a numerical solution
of Eq.\ \ref{EqSE}. \ The limiting expression for $\tau \rightarrow 1$ gives
the standard Landau-Zener formula appropriate for a two-state system. \ In
the tunnel-junction limit $\tau \rightarrow 0$ one finds%
\begin{equation}
W=(\tau /2)^{2\Delta /eV}\text{ .}  \label{Eqphoton}
\end{equation}

\FRAME{ftbpFU}{4.3439in}{2.2771in}{0pt}{\Qcb{(a) Plot of average junction
current $\left\langle I_{j}\right\rangle $ versus inverse DC voltage $V$ for
transmission coefficients $\protect\tau =0.8$, $0.5$, $0.2$, and $0.01$
(from Ref. \protect\cite{Bratus}\ ). \ Solid lines are from exact
calculation, and dashed lines are from predictons of Eqs. \protect\ref{EqIqp}
and \protect\ref{EqSE}. The time dependence of the Josephson current $I_{J}$
is plotted for the ground state (b) and for a transition (c), where the
insets show the trajectory of the bound states as $E_{J}$ \textit{vs}. $%
\protect\delta $. \ \ }}{\Qlb{figCompare}}{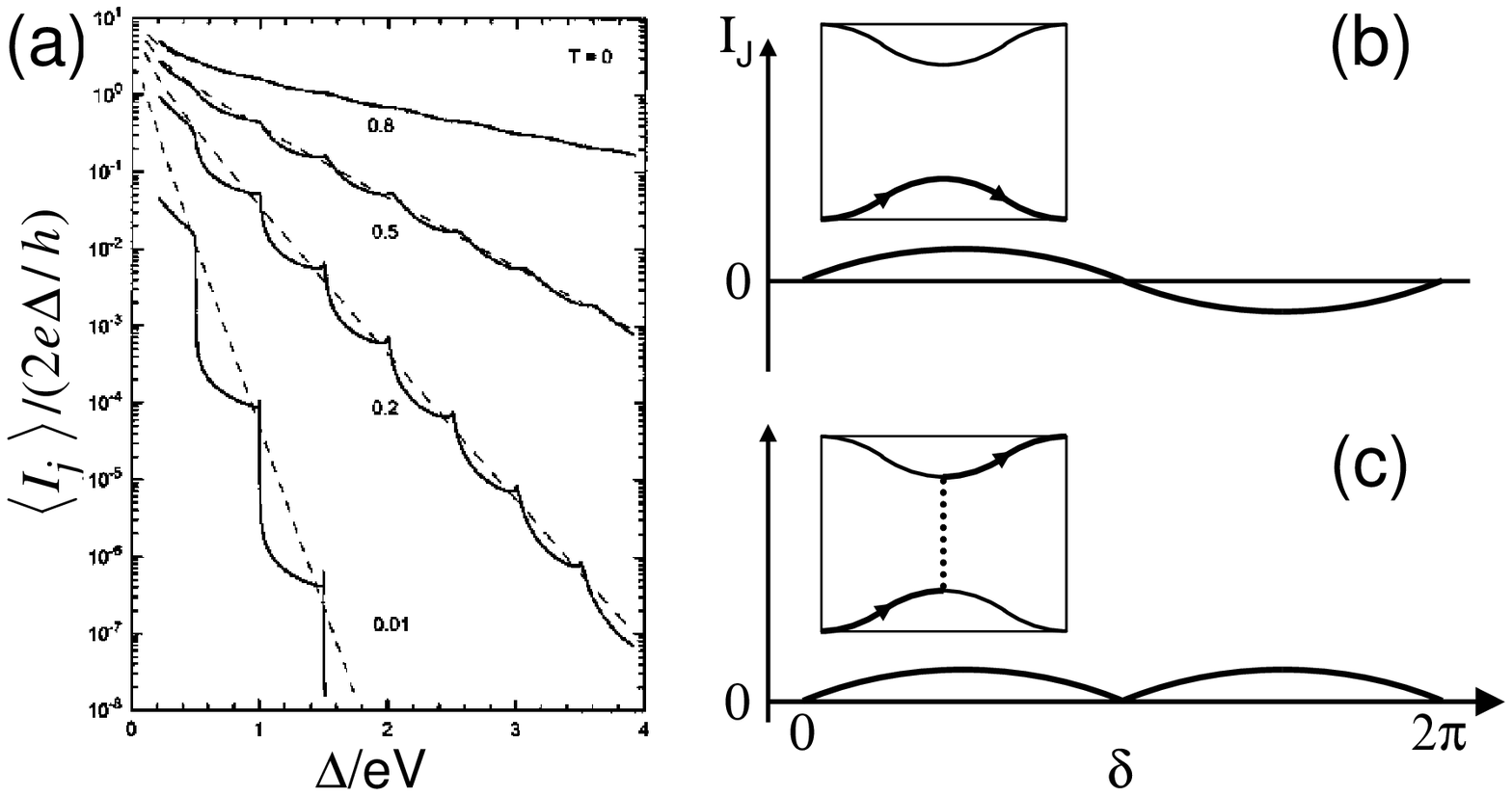}{\special{language
"Scientific Word";type "GRAPHIC";maintain-aspect-ratio TRUE;display
"USEDEF";valid_file "F";width 4.3439in;height 2.2771in;depth
0pt;original-width 5.7597in;original-height 7.3613in;cropleft "0";croptop
"1";cropright "0.9996";cropbottom "0";filename 'fig11.eps';file-properties
"XNPEU";}}

For the case of a constant DC bias voltage $V$, the total junction current $%
\left\langle I_{j}\right\rangle $ may be calculated with this transition
rate and an attempt rate $\Gamma =(2e/h)V$ given by the frequency at which $%
\delta $ passes $\pi /2$. \ Using Eq. \ref{EqSE} and setting the power of
quasiparticle generation $2\Delta \Gamma W$ to the electrical power $%
\left\langle I_{j}\right\rangle V$, one finds 
\begin{equation}
\left\langle I_{j}\right\rangle =\frac{2e\Delta }{\pi \hbar }W\text{ .}
\label{EqIqp}
\end{equation}%
This prediction is plotted in Fig. \ref{figCompare}(a) and shows very good
agreement with the results of exact calculations\cite{Bratus}. \ Only the
steps in voltage are not reproduced, which are understood as arising from
the quantization of energy $eV$ from multiple Andreev reflection of the
quasiparticles. \ The steps are not expected to be reproduced by the
semiclassical theory since this theory is an expansion around small
voltages, or equivalently, large quantization numbers. \ 

The junction current may also be determined from the energies of the two
bound states. \ For a constant voltage across the junction, we use Eq. \ref%
{UtoI} to calculate the charge transferred across the junction after a phase
change of $2\pi $ 
\begin{eqnarray}
Q_{j} &=&\int_{0}^{2\pi /(d\delta /dt)}I_{j}\,dt  \label{EqQja} \\
&=&\frac{2\pi }{\Phi _{0}}\int_{0}^{2\pi /(d\delta /dt)}\frac{dU_{J}}{%
d\delta }dt  \label{EqQjb} \\
&=&\frac{\left[ U_{J}(2\pi )-U_{J}(0)\right] }{V}\text{ ,}  \label{EqQjc}
\end{eqnarray}%
which gives the expected result that the change of energy equals $Q_{j}V$. \
When the junction remains in the ground state, the energy is constant $%
U_{J}(2\pi )-U_{J}(0)=0$ and no net charge flows through the junction. \ Net
charge is transferred, however, after a transition. \ The charge transfer $%
2\Delta /V$ multiplied by the transition rate gives an average current $%
Q_{j}\Gamma W$ that is equivalent to Eq. \ref{EqIqp}. \ 

Equation \ref{EqIj} may be used to calculate the time dependence of the
Josephson current, as illustrated in Fig. \ref{figCompare}(b) and (c). \
When the system remains in the ground state (b), the junction current is
sinusoidal and averages to zero. \ For the case of a transition (c), the
current before the transition is the same, but the Josephson current remains
positive after the transition (see Eq. 5.4 of Ref. \cite{Bratus}). \ The
transition itself also produces charge transfer from multiple-Andreev
reflections(MAR) \cite{Averin,MAR} 
\begin{equation}
Q_{\mathrm{MAR}}=2\Delta (1-\tau )^{1/2}/V\text{ .}
\end{equation}%
This result is perhaps surprising - the junction current at finite voltage
arises from transfer of charge $Q_{\mathrm{MAR}}$ \textit{and} a change in
the Josephson current. \ The relative contribution of these two currents is
determined by the relative size of the gap in the bound states. \ For $\tau
\rightarrow 1$\ , all of the junction current is produced by Josephson
current, whereas for $\tau \rightarrow 0$ (tunnel junctions)\ the current
comes from $Q_{\mathrm{MAR}}$. \ 

For small voltages, the transition event must transfer a large amount of
charge $Q_{\mathrm{MAR}}$ in order to overcome the energy gap. \ In
comparing this semi-classical theory with the exact MAR theory, $Q_{\mathrm{%
MAR}}/e$ has an integer value and represents the order of the MAR process
and the number of electrons that are transferred in the transition. \ This
description is consistent with Eq. \ref{Eqphoton} describing the transition
probability for an $n$-th order MAR process, where $n=2\Delta /eV$, and $%
\tau /2$ represents the matrix element for each order. \ 

\ From this example it is clearly incorrect to picture the quasiparticle and
Josephson current as separate entities, as suggested by the calculations of
perturbation theory. \ To do so ignores the fact that quasiparticle
tunneling, arising from a transition between the bound states, also changes
the Josephson contribution to the current from $\delta =\pi $ to $2\pi $. \ 

\section{\label{QPandQubits}Quasiparticle Bound States and Qubit Coherence}

The quasiparticle bound-state theory can be used to predict both the
Josephson and quasiparticle current in the zero-voltage state, as
appropriate for qubits. \ In this theory an excitation from the $E_{J-}$
bound state to the $E_{J+}$ state is clearly deleterious as it will change
the Josephson current, fluctuating the qubit frequency and producing
decoherence in the phase of the qubit state. \ For an excitation in one
channel, the fractional change in the Josephson current is $\sim 1/N_{%
\mathrm{ch}}$, where $N_{\mathrm{ch}}$ is the number of conduction channels.
\ The subgap current-voltage characteristics can be used to estimate $N_{%
\mathrm{ch}}$, which gives an areal density of $\sim 10^{4}/\mu \mathrm{m}%
^{2}$ \cite{Simmonds,Lang}. \ For a charge qubit with junction area $%
10^{-2}\mu \mathrm{m}^{2}$, the qubit frequency changes fractionally by $%
\sim 1/N_{\mathrm{ch}}\sim 10^{-2}$ for a single excitation, and gives
strong decoherence. \ Although the phase qubit has a smaller change $(1/N_{%
\mathrm{ch}})I_{0}/4(I_{0}-I)\sim 2\times 10^{-5}$, the excitation of even a
single bound state is clearly unwanted. \ 

Fortunately, these quasiparticle bound states should not be excited in
tunnel junctions by the dynamical behavior of the qubit. \ The $E_{J-}$ to $%
E_{J+}$ transition is energetically forbidden because the energy of the
qubit states are typically choosen to be much less than $2\Delta $. \ Thus,
the energy gap of the superconductor protects the qubit from quasiparticle
decoherence.

If a junction has \textquotedblleft pinhole\textquotedblright\ defects,
where a few channels have $\tau \rightarrow 1$, then the energy gap will
shrink to zero at $\delta =\pi $. \ However, only the flux qubit will be
sensitive to quasiparticles produced at these defects since it operates near 
$\delta =\pi $. \ In contrast, the phase qubit always retains an energy gap
of at least $\sqrt{2}\Delta $ around its operating point $\delta =\pi /2$
(see arrow in Fig. \ref{figsBS}). \ We note this idea implies that a phase
qubit can even be constructed from a microbridge junction, which has some
channels\cite{Dorokov} with $\tau =1$. \ Although the phase qubit is
completely insensitive to pinhole defects, this advantage is probably
unimportant because Al-based tunnel junctions have oxide barriers of good
quality. \ 

Pinhole defects also change the Josephson potential away from the -$\cos
\delta $ form. \ This modification is typically unimportant because the
deviation is smooth and can be accounted for by a small effective change in
the critical current. \ 

The concept that the energy gap $\Delta $ protects the junction from
quasiparticle transitions suggests that superconductors with nonuniform gaps
may not be suitable for qubits. \ Besides the obvious problem of conduction
channels with zero gap, channels with a reduced gap may cause stray
quasiparticles to be trapped at the junction. \ The high-$T_{c}$
superconductors, with the gap suppressed to zero at certain crystal angles,
are an obvious undesirable candidate. \ However, even Nb could be
problematic since it has several oxides that have reduced or even zero gap.
\ Nb based tri-layers may also be undesirable since the thin Al layer near
the junction slightly reduces the gap around the junction. \ In contrast, Al
may not have this difficulty since its gap \textit{increases} with the
incorporation of oxygen or other scattering defects. \ It is possible that
these ideas explain why Nb-based qubits do not have coherence times as long
as Al qubits\cite{Martinis,Simmonds}.

\section{\label{summary}Summary}

In summary, Josephson qubits are nonlinear resonators whose critical element
is the nonlinear inductance of the Josephson junction. \ The three types of
superconducting qubits, phase, flux, and charge, use this nonlinearity
differently and produce qubit states from a cubic, quartic, and cosine
potential, respectively.\ 

To understand the origin and properties of the Josephson effect, we have
first reviewed the BCS theory of superconductivity. \ The superconducting
phase was explicitly shown to be a macroscopic property of the
superconductor, whose classical and quantum behavior is determined by the
external electrical circuit. \ After a review of quasiparticle and Josephson
tunneling, we argued that a proper microscopic understanding of the junction
could arise only from an exact solution of the Josephson effect.

This exact solution was derived by use of mesoscopic theory and
quasiparticle bound states, where we showed that Josephson and quasiparticle
tunneling can be understood from the energy of the bound states and their
transitions, respectively. \ A semiclassical theory was used to calculate
the transition rate for a finite DC voltage, with the predictions matching
well that obtained from exact methods. \ \ 

This picture of the Josephson junction allows a proper understanding of the
Josephson qubit state. \ We argue that the gap of the superconductor
strongly protects the junction from quasiparticle tunneling and its
decoherence. \ We caution that an improper choice of materials might give
decoherence from quasiparticles that are trapped at sites near the junction.

We believe a key to future success is understanding and improving this
remarkable nonlinearity of the Josephson inductance. \ We hope that the
picture given here of the Josephson effect will help researchers in their
quest to make better superconducting qubits. \ 

\section{Acknowledgements}

We thank C. Urbina, D. Esteve, M. Devoret, and V. Shumeiko for helpful
discussions. \ This work is supported in part by the NSA\ under contract
MOD709001.\


\begin{thebibliography}{99}
\bibitem{states} Our notation for incoming and outgoing states is choosen to
correspond to the boundry conditions for scatttering in Ref.\ \cite{Bratus}.
\end{thebibliography}
\end{document}